\documentclass[%
reprint,
superscriptaddress,
showpacs,
amsmath,amssymb,
aps,
pra,
floatfix,
a4paper,
twocolumn
]{revtex4-1}

\usepackage{graphicx}% Include figure files
\usepackage{dcolumn}% Align table columns on decimal point
\usepackage{bm}% bold math
\usepackage{mathrsfs}
\usepackage{color}
\usepackage{mathtools}
\usepackage{amsmath}
\usepackage{mhchem}
\usepackage{tensor}
\usepackage{float}
\newcommand{\unit}{1\hspace{-0.24em}\mathrm{l}}

\def\ket#1{|#1\rangle}
\def\bnu{{\mathbf{\nu}}}

\def\openone{{\mathchoice {\rm 1\mskip-4mu l}   {\rm 1\mskip-4mu l}
                      {\rm 1\mskip-4.5mu l} {\rm 1\mskip-5mu l}}}
\let\id\openone

\begin{document}
\title{Multirobust control pulses designed by \\
	iterative quantum Kolmogorov-Arnold-Moser algorithm}
\author{M. Harutyunyan}
\affiliation{Universit\'e Bourgogne Europe, Laboratoire Interdisciplinaire Carnot de Bourgogne ICB, CNRS UMR 6303, 
	21000 Dijon, France}
\author{S. Gu\'erin}
\email{sguerin@ube.fr}
\affiliation{Universit\'e Bourgogne Europe, Laboratoire Interdisciplinaire Carnot de Bourgogne ICB, CNRS UMR 6303,
	21000 Dijon, France}

\date{\today}
\begin{abstract}
	We introduce a multi-robust quantum control strategy designed to suppress multilevel leakage outside the computational qubit space while simultaneously ensuring robustness against unknown inhomogeneities in control parameters. It is based on a reformulation of the Derivative Removal by Adiabatic Gate (DRAG) technique which allows for the full incorporation of control pulses originally designed for two-level models by using the iterative quantum Kolmogorov-Arnold-Moser (KAM) algorithm. We apply it for optimal robust control designed by inverse optimization (RIO), referred to as RDK pulses.
	We analyze the performance of RDK pulses on various durations, iterations, and robustness types. We show that the two-level robustness is restored after only a few iterations 
	with an asymptotic exponential approximation (i.e. as a superasymptotic series), even for relatively short pulse durations and that the resulting pulses are near optimal in terms of energy compared to the optimal RIO pulses defined in the qubit subspace. Such pulses are well suited for superconducting transmons as well as other forms of qubits embedded in a multi-level structure. By effectively reproducing a two-level dynamics within a multi-level system, this approach is versatile thanks to its compatibility with two-level control strategies under conditions (smoothness, duration) that are analyzed. 
\end{abstract}
\maketitle

\textit{Introduction.} High-fidelity quantum gate operations are crucial for a wide range of quantum applications, particularly quantum computing \cite{Nielsen, Koch}.
Techniques of quantum control that are in addition resilient against errors, such as deviations of the controls or imperfections of the model, have significantly improved gate performance. One can mention adiabatic passage \cite{SWAP,Vitanov_Raman,Sorin}, shortcut to adiabaticity \cite{STA,robustNJP,SSSP,Clerk,STAAQC}, composite pulse techniques \cite{Wimperis,Cummins,Torosov,Jones,Torosov2,Gevorgyan}, the use of geometric methods \cite{Barnes} including robust inverse optimization (RIO) \cite{Dridi, Laforgue,Laforgue2}, and the derivative removal by adiabatic gate (DRAG) to eliminate population leakage outside the qubit subspace \cite{Motzoi2009,Gambetta2011,Theis2018,Werninghaus2021}. The latter is particularly crucial for the widely used superconducting qubits for which their weak anharmonicity makes them prone to such leakage. 
However, simultaneous protection against these diverse sources of errors, specifically control inhomogeneities \textit{and} leakage, is a persistent challenge. Non-optimal solutions based on restrictive pulse ansatz invoking either composite pulses \cite{Tonchev} or Fourier-type series \cite{YongHai} have been proposed and implemented \cite{KangyuanYi}. 

In this Letter, we propose an algorithm that combines and generalizes the principle of DRAG, implemented by the quantum analogue of the Kolmogorov-Arnold-Moser (KAM) perturbation theory \cite{Scherer-II-97,Scherer-prl95,PhysicA,Review}, with optimal control techniques robust with respect to unknown inhomogeneities of the controls. Despite the apparent complexity of the multirobust control problem, the proposed technique offers an optimal control without pulse ansatz nor gradient-based numerical procedure. The KAM technique offers a way to construct a sequence of transformations that systematically lower the order of non-diagonal terms of the Hamiltonian. This combined approach generates thus control protocols that systematically reduce the influence of higher-energy states by eliminating the terms coupling them to the qubit subspace while simultaneously ensuring that the original controls within the qubit subspace remain unaltered. In fact, this technique remarkably allows the compatibility with any two-level control strategy. In this Letter, we implement it using RIO pulses \cite{Dridi, Laforgue,Laforgue2}. The combined technique RIO with the DRAG-KAM algorithm, leading to pulses referred to as RDK pulses, is thus shown to offer multi-robust pulses. We exhibit the effect of the iterative DRAG-KAM technique on the resonant $\pi$-pulse, as well as on two types of RIO pulses: those robust against pulse amplitude inhomogeneities ($\alpha$-RIO), and those robust against both amplitude and detuning inhomogeneities ($\alpha\delta$-RIO).

\textit{KAM technique.}
The KAM technique is a perturbative approach based on iterative unitary transformations that reduce coupling terms \cite{Scherer-II-97,Scherer-prl95,PhysicA,Review}. By identifying a relevant subspace in Hilbert space (here the qubit subspace), the method constructs an effective Hamiltonian that minimizes interactions with the complementary subspace, enabling a simplified description of the dynamics.

We decompose the Hamiltonian with an unperturbed Hamiltonian
$H_{0}$ and a perturbation $\varepsilon V_1$ of order $\varepsilon$:
\begin{equation}
	H=H_{0}+\varepsilon V_1.  \label{K1}
\end{equation}
The
method is defined quite generally, independently of the particular
form of $H_{0}$ and $V_1$. We assume that $V_1$ is a bounded
operator. It consists in constructing a unitary transformation
$e^{\varepsilon W_{1}}$, with $W_{1}^{\dag }=-W_{1}$ such that
\begin{equation}
	e^{-\varepsilon W_{1}}He^{\varepsilon W_{1}}=H_{0,1}+\varepsilon ^{2}V_{2}=:H_{2}, \ H_{0,1}=H_{0}+\varepsilon
	D_{1},  \label{epep2}
\end{equation}
where $D_{1}$ is a diagonal (or block-diagonal for the considered case of the qubit subspace) operator with respect to $H_0$, i.e. satisfying $%
[H_{0},D_{1}]=0 $. Thus the perturbation is reduced from order $%
\varepsilon $ to $\varepsilon ^{2}$.

An equation to determine $D_{1}$
and $W_{1}$ can be obtained by expanding Eq. (\ref{epep2}) in powers
of $\varepsilon $ and requiring that the terms
of order $\varepsilon $ cancel out \cite{Review}. From the resulting cohomology equations: $D_{1} :=[H_{0},W_{1}]+V_{1}$ and
$[H_{0},D_{1}] =0$, $D_1$ and $W_1$ can be determined in terms of the eigenvalues
and eigenvectors of $H_{0}$ denoted as $\lambda^0_{\bnu}$ and $%
\ket{\bnu,j}$, respectively (the index $\bnu$ labels the
different eigenvalues, and $j$ distinguishes different basis
vectors corresponding to a degenerate eigenvalue):
\begin{subequations}
	\label{sol_coh}
	\begin{eqnarray}
		D_{1} &=&\sum_{\bnu,j,j^{\prime }}|\bnu,j\rangle \langle \bnu%
		,j|V_{1}|\bnu,j^{\prime }\rangle \langle \bnu,j^{\prime }|,  \label{solWD} \\
		W_{1} &=&-\sum_{\bnu,j,j^{\prime },\bnu^{\prime }\not=\bnu}{\frac{|\bnu%
				,j\rangle \langle \bnu,j|V_1|\bnu^{\prime },j^{\prime }\rangle \langle \bnu%
				^{\prime },j^{\prime }|}{\lambda _{\bnu}^{0}-\lambda
				_{\bnu^{\prime }}^{0}}}.
		\label{solWD_}
	\end{eqnarray}
\end{subequations}
The solution $W_{1}$ is not unique, since if $A$ is any operator such that $%
[H_{0},A]=0$ then $W_{1}+A$ is also a solution. This allows free parameters in the definition of the $W_1$, which will be used to accelerate the convergence.

Since the transformed Hamiltonian $H_2$ retains the same general form as $H$, the procedure can be iterated by defining appropriate $H_{0,j}$ and $W_j$. Each step reduces the perturbation from order $\varepsilon$ to $\varepsilon^2$, then to $\varepsilon^4$, and so on, yielding a superexponential decrease of order $\varepsilon^{2^N}$ at $N$-th iteration. It is referred to as a quantum KAM algorithm, in analogy with the classical KAM transformations \cite{Scherer-II-97},  and the transformations $e^{\varepsilon^{2^{N-1}} W_{N}}$ are called KAM transformations (or contact transformations). 
\textcolor{black}{Under some particular conditions, typically small enough $\varepsilon$ and excluding resonances from small denominators in the $W_j$'s \cite{PhysicA}, as it is done for the classical invariant KAM tori, the iterations above show in principle superexponential convergence as a function of $N$. }

Here we use a simpler, non-superexponential
version, keeping the same $H_0$ at each iteration, which already
gives remarkable exponential convergence in $\varepsilon$ for
an optimal iteration number.
A series expansion indeed behaves in general as an asymptotic series as a function of $\varepsilon$ for fixed $N$, which is divergent if one considers the limit of a large $N$ for a finite $\varepsilon$ \cite{Bender}. If one makes only a finite number $N$ of iterations, choosing $N$ optimally, one can obtain an asymptotic formula with an error that is exponentially small in $\varepsilon$,  i.e. of the form $e^{-C/\varepsilon}$ (with $C$ a positive constant), which is referred to  as
{\it superasymptotic series}   \cite{BerryHyper,Boyd}. 
This is similar to superadiabatic theory \cite{BerrySA,Joye,Holthaus}. We show here the emergence of such exponential
asymptotics in KAM-based leakage suppression for qubit control.

In time-dependent settings, the transformation of the Schr\"odinger equation introduces an additional dynamical time-derivative perturbative term, $-i\, e^{-\varepsilon W_1} \frac{d}{dt} e^{\varepsilon W_1}$. We notice that it reduces to $-i \varepsilon \dot{W}_1$ if the time dependence of $W_1$ factorizes such that $[\dot{W}_1, W_1] = 0$.

\textit{DRAG technique.}
A quantum system often consists of a two‑level computational (qubit) subspace embedded within a larger Hilbert space containing leakage states. The total Hamiltonian can be modeled as
$  H(t) = H_0 + \varepsilon V(t)$,
where the small parameter $\varepsilon$ allows to identify the size of the perturbation, which in general contains all the orders:
$ \varepsilon V(t) = \sum_{N=1}^{\infty} \varepsilon^N V^{(N)}(t)$.
Here $V^{(1)}(t)$ is the controls (amplitude and phase/detuning) we wish to apply to the qubit, which determine the original control strategy in the qubit subspace.

The goal of DRAG is to find a time dependent unitary transformation $A(t)$
such that the effective Hamiltonian in the new frame,
$ H_{\rm eff}(t) = A^\dagger(t) H(t) A(t) - i\,A^\dagger(t)\dot A(t)$,
is block-diagonal between the qubit and leakage subspaces \cite{Motzoi2009,Gambetta2011,Theis2018,Werninghaus2021}. This removes all off-diagonal couplings to the desired order in $\varepsilon$, ensuring that any state initialized in the qubit manifold remains there throughout the gate. The transformation should not affect the boundary states at the beginning and end of the control, therefore $A(t_i)=A(t_f)=\unit$ at the initial $t_i$ and final $t_f$ times, respectively.

DRAG method has been originally designed for implementing a resonant $\pi$-pulse control with, as the leading order correction, the compensation of the derivative of the control field amplitude. However, we show here that it can be formulated in a more general setting: by enforcing the qubit block of $H_{\rm eff}(t)$ to fully match $V^{(1)}(t)$, we make DRAG become a method for designing controls that replicate those of the qubit-subspace control.
In this Letter, we reformulate and implement the DRAG method using the iterative quantum KAM algorithm, i.e. the DRAG-KAM technique, such that it allows for the full incorporation of control pulses originally designed for two-level models, the RIO pulses.

\textit{Iterative DRAG-KAM algorithm.}
For the sake of simplicity of presentation, we consider the problem with the three lowest states $\vert 0\rangle,\vert 1\rangle,\vert 2\rangle$, but the algorithm can be generalized for any multilevel problem. The Hamiltonian reads: 
\begin{equation}
	\label{DefHam}
	H= \begin{pmatrix}0 & \bar\Omega/2 &  0 \cr 
		\Omega/2 & \Delta & \lambda \bar\Omega/2  \cr
		0& \lambda\Omega/2  & d + 2\Delta
	\end{pmatrix} \equiv  \begin{pmatrix} H_{00} & V_{01} \cr 
		V_{01}^{\dagger} & d + 2\Delta
	\end{pmatrix}
\end{equation}
with the driving field of Rabi frequency amplitude $\Omega=\Omega_x+i\Omega_y$, $\Delta$ the corresponding detuning, and $d$ the anharmonicity (which can be positive or negative). For a weakly anharmonic oscillator $\lambda=\sqrt{2}$. The goal is to eliminate dynamically the effect of the upper state $\vert 2\rangle$ and thus of the unwanted coupling $\lambda\Omega$. \textcolor{black}{The Hamiltonian \eqref{DefHam} has been presented in two coupled blocks with $H_{00}$ featuring the dynamics in the qubit space. This coupling features two effects that have to be corrected: an effective leakage of the population to state $\vert 2\rangle$ and a modification of the original gate. This is formally achieved by block-diagonalizing \eqref{DefHam} by a tranformation $T_1$ and incorporating the resulting dynamical term:  
	\begin{equation}
		T_1^{\dagger} H T_1 -i T_1^{\dagger} \dot T_1=  \begin{pmatrix} H_{1,00} & V_{1,01} \cr 
			V^{\dagger}_{1,01} & H_{1,11}
		\end{pmatrix},
	\end{equation}
	where $V_{1,01}$ includes solely block-off-diagonal terms of $-i T_1^{\dagger} \dot T_1$ and $H_{1,jj}$, $j=0,1$ the block-eigenvalues from $T_1^{\dagger} H T_1$ and block-diagonal elements from $-i T_1^{\dagger} \dot T_1$. The corrections are constructed such that an additional coupling compensates both $V_{1,01}$ to prevent the leakage and $H_{1,00}$ in order to recover the original gate dynamics $H_{00}$. This program is implemented by the iterative DRAG-KAM transformations which is detailed in the End Matter. The code used to produce the data in this article is publicly available \cite{code-data}.
	The procedure is reminiscent to a superadiabatic transformation $T_1$ combined with a counterdiabatic driving compensating the non-adiabatic coupling $V_{1,01}$ \cite{Unanyan,Berry,XiChen}.}

\textit{RIO technique.}
Variation or imperfect knowledge of the pulse parameters can lead to loss of fidelity. RIO has been proposed to address this issue by optimally solving the control problem in the qubit space using Euler-Lagrange optimization with the Single-Shot Shaped Pulse method \cite{Dridi, Laforgue}. The Hamiltonian in the qubit space $\{\vert 0\rangle,\vert 1\rangle\}$ is given by:
\begin{equation}
	H =  \frac{1}{2}\begin{pmatrix}-\Delta & \bar\Omega  \cr 
		\Omega & \Delta  \cr
	\end{pmatrix} + \frac{1}{2}\begin{pmatrix} -\delta  & \alpha\bar\Omega  \cr 
		\alpha\Omega & \delta  \cr
	\end{pmatrix},
\end{equation}
where the relative parameter $\alpha$ accounts for static field inhomogeneities and $\delta$ represents the so-called inhomogeneous broadening or quasi-static energy-level noise. Solutions have been developed for robustness against $\alpha$ alone, and can be applied to simultaneous robustness against $\alpha$ and $\delta$. However, for these solutions, $\Omega(t)$ does not vanish at the boundaries of the pulse, which is an essential requirement for compatibility with the iterative DRAG-KAM algorithm discussed above. To resolve this, we introduce an additional ingredient in the construction: \textit{a time reparametrization procedure}, which preserves the pulse area \cite{ZhuJPB}. It is shown in the End Matter.

\textit{Numerical results.}
We show the efficiency of the iterative DRAG-KAM algorithm for three types of pulses: resonant $\pi$-pulse, $\alpha$-RIO and $\alpha\delta$-RIO. Coupled with the DRAG-KAM algorithm, the $\alpha\delta$-RIO pulse is designed to achieve near-optimal multi-robustness, simultaneously addressing pulse inhomogeneities, variations in detuning, and state leakage.

Throughout our analysis, we consider complete population transfer from the initial state $|\phi(t_i)\rangle = \lvert 0 \rangle$ to the target state $| \phi_T\rangle=| 1\rangle$ up to a phase, easily generalized to other partial
transfers and qubit gates. The infidelity is defined as $\log_{10}(1 - |  \langle \phi_T | \phi(t_f)\rangle|^2)$. 
We ensure the transformation $T_1$ reduces to $\openone$ at the initial and final
times by imposing $\Omega^{(1)}(t_i)=  \Omega^{(1)}(t_f)= 0$, satisfied for
Gaussian-shaped pulses.

\begin{figure}[h]
	\centering
	\includegraphics[width=0.9\linewidth]{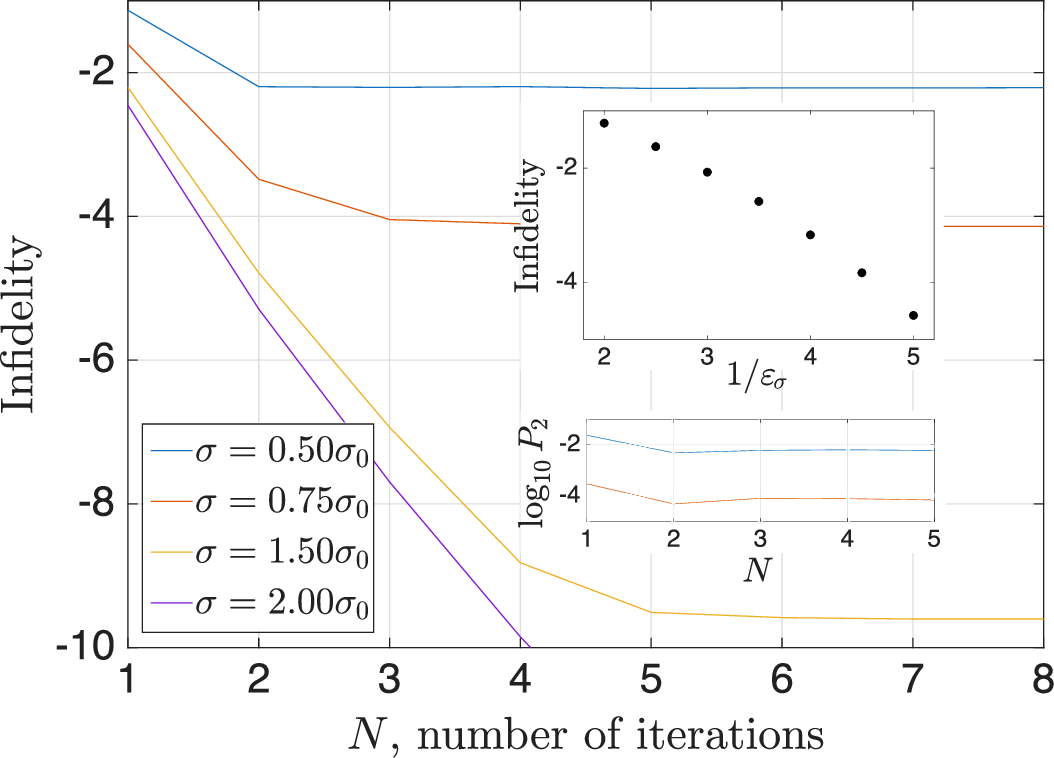}
	\caption{Infidelity of complete population transfer, $\log_{10}(1 - |\langle1|\phi(t_f)\rangle|^2)$, driven by a resonant $\pi$-pulse for various pulse durations up to $N=8$ iterations, under the iterative DRAG-KAM algorithm. We observe an infidelity of the order of $-4$, compliant to quantum technologies applications, for pulses as short as $\sigma = 0.75\sigma_0$ (with the normalization $\sigma_0=2\pi/\vert d\vert$ and a negative anharmonicity $d=-\vert d \vert$).  \textcolor{black}{The lower inset shows the leakage population $P_2= \vert \langle 2 | \phi(t_f)\rangle|^2$ into the upper state (in logarithmic scale), which reaches its minimum after a single iteration $N=2$ thanks to the optimal $\kappa=0.5$. The upper inset displays the minimum infidelity as a function of $1/\varepsilon_{\sigma}\equiv \sigma \vert d\vert$ in logarithmic scale showing its superasymptotic exponential form $e^{-C/\varepsilon_{\sigma}}$ for small enough $\varepsilon_{\sigma}<1/2$. }}
	\label{fig:DRAG_pi}
\end{figure}

\begin{figure}[h]
	\centering
	\includegraphics[width=0.9\linewidth]{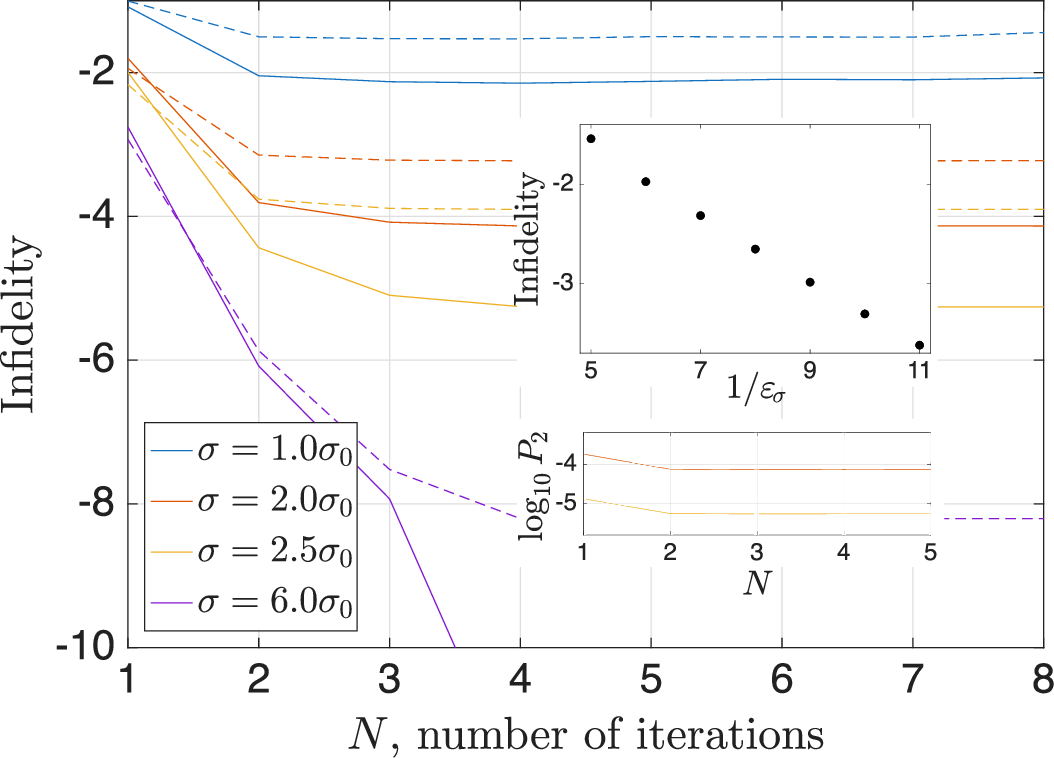}
	\caption{Same as Fig. \ref{fig:DRAG_pi} but for $\alpha$-RIO (solid lines) and $\alpha\delta$-RIO (dashed lines) pulses. The infidelity $-4$ is reached for $\sigma=2\sigma_0$ ($\sigma=2.5\sigma_0$) for $\alpha$-RIO ($\alpha\delta$-RIO). The insets are for $\alpha$-RIO.}
	\label{fig:RIOs}
\end{figure}

We begin by examining the simplest case: a resonant $\pi$-pulse, characterized by $\Delta^{(1)} = 0$ and $\int_{t_i}^{t_f} \Omega^{(1)}(t) \, dt = \pi$. The infidelity over successive iterations up to $N = 8$ is shown in Fig.~\ref{fig:DRAG_pi} for various durations. 
We define $N = 1$ as the baseline iteration, before any leakagemitigating
corrections. Since the pulse area is fixed, increasing $\sigma$ decreases $\max(\Omega^{(1)}(t))/|d|$, reducing the third
level's influence and easing leakage suppression. The DRAG-KAM defines a free parameter $\kappa$ that can be chosen for optimizing the reduction of the leakage and the fidelity (see End Matter).
Here and in all subsequent cases, we take $\kappa = 0.5$ as the best fidelity-robustness compromise, $\kappa = 0$ giving the best robustness while $\kappa\approx 1$ gives lower infidelities
at its expense. The fidelity improves over successive 4 iterations before plateauing after a few steps; the number
of iterations to reach the plateau grows with $\sigma$, but so does the achievable fidelity.
\textcolor{black}{We have determined numerically that the resulting solution behaves as a superasymptotic series with an error of exponential form $e^{-C/\varepsilon_{\sigma}}$ (with $C$ a positive constant equal 2.5 for the $\pi$-pulse and 0.7 for $\alpha$-RIO) where we define here $\varepsilon_{\sigma}$ as $\varepsilon_{\sigma} = 1/(\sigma \vert d\vert)$,  which is proportional to the static parameter $\varepsilon$, because for the pulses discussed in this paper $\varepsilon = \max(|\Omega|,|\Delta|)/|d| \equiv \max(|\Omega|)/|d|$ and $\max(|\Omega|)\sim 1/\sigma$ due to Eq.~\eqref{eq:timereparsig}. This is shown in the upper inset of Fig. \ref{fig:DRAG_pi}. 
	We have analyzed how the corrections are achieved in the algorithm. The lower inset of Fig. \ref{fig:DRAG_pi} shows that the leakage population into the upper state reaches its minimum value in a single iteration thanks to the optimization of $\kappa$. Further iterations allow to correct the transfer while preserving the minimum leakage. This analysis also applies for RIO pulses as shown in Fig. \ref{fig:RIOs}.}

Ultimately, for sufficiently long pulses, we achieve machine-level accuracy. 
For relatively short pulses, such as $\sigma = 0.75 \sigma_0$, an infidelity as low as $-4$ is attainable, corresponding to a typical ultra-high fidelity required for quantum computing applications. 
When the pulses are too short, the algorithm does not improve the fidelity for $N>2$. 
In addition, we have examined the impact of eliminating leakage to the fourth and fifth levels in the presence of a multilevel anharmonic ladder. These contributions were found to be negligible. The sign of $d$ also does not have any significant effect on the results of the resonant $\pi$-pulse.

\begin{figure}[h]
	\centering
	\includegraphics[width=0.9\linewidth]{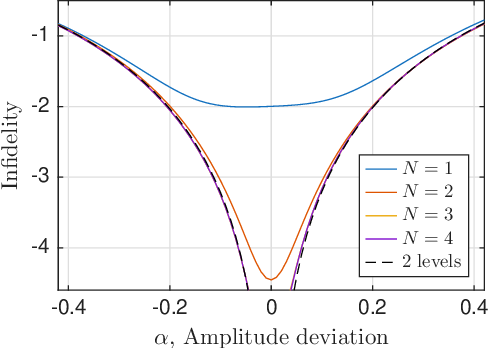}
	\caption{Infidelity profile versus $\alpha$ (dimensionless)  at each iteration up to 4 iterations for $\alpha$-RIO with $\sigma = 2.5 \sigma_0$. DRAG-KAM recovers the robustness of two-level RIO controls.}
	\label{fig:Robustness_alphario_alpha}
\end{figure}

Now we consider the $\alpha$-RIO pulse which is robust to variations in $\alpha$, and the $\alpha\delta$-RIO pulse which maintains robustness against changes in both $\alpha$ and $\delta$.
As both $\alpha$-RIO and $\alpha\delta$-RIO pulses do not satisfy the condition $\Omega^{(1)}(t_i) = \Omega^{(1)}(t_f) = 0$ \cite{Dridi, Laforgue,Laforgue2}, we apply the time reparametrization technique described in the End Matter before applying the iterative DRAG-KAM algorithm. The results are presented in Fig.~\ref{fig:RIOs}.
A nonzero detuning increases the effects of higher levels, making them more difficult to suppress. As a result, larger values of $\sigma$ are required to achieve the same infidelities as those obtained with the resonant pulse (Fig. 1).  For $\alpha$-RIO ($\alpha\delta$-RIO) an infidelity of $-4$ is achieved for pulses as short as $\sigma = 2\sigma_0$ ($\sigma = 2.5\sigma_0$).
\textcolor{black}{We have found that the algorithm works for values of $\varepsilon$ as large as approximately $0.5$. In particular, the increase in infidelity observed 
	for the short $\alpha\delta$-RIO pulse originates from the detuning 
	corrections, which effectively  increase $\varepsilon$ compared to the $\pi$-pulse.}

Figure ~\ref{fig:Robustness_alphario_alpha} shows the resulting robustness profiles after various iterations of $\alpha$-RIO pulse with $\sigma = 2.5\sigma_0$. 
The $\alpha$-profile for $\alpha\delta$-RIO pulse (not shown) is very similar to the one shown in Fig.~\ref{fig:Robustness_alphario_alpha}.  The $\delta$-profile is shown in Fig. 5 in the End Matter. We observe that DRAG-KAM recovers the RIO robustness determined in the space restricted to the two-level after three iterations. 
We have found that, unlike for the $\pi$-pulse, the fidelities obtained depend slightly on the sign of $d$.

An important feature is that the application of the DRAG-KAM algorithm induces RDK pulses with only small modifications in the optimal pulse area and energy of RIO (typically $10\%$ with an hyper-Gaussian envelope and less than $5\%$ for a longer Gaussian pulse). We can thus conclude that the resulting multi-robustness is near optimal (compared to the absolute optimum in the case of $\alpha$-RIO and $\alpha\delta$-RIO determined in the space restricted to the one of the qubit).

\textit{Conclusion and discussion.}
We have introduced an iterative algorithm that combines the DRAG technique with the iterative KAM method, enabling the design of control fields within realistic multi-level quantum systems that effectively replicate two-level control protocols. 
In particular, we have demonstrated its compatibility with both $\alpha$-RIO and $\alpha\delta$-RIO pulses in the case of complete population transfer, but the technique applies to other types of transfer or gates as well. The method achieves thus multi-robustness, simultaneously addressing detuning variations, amplitude inhomogeneities, and state leakage while being near optimal in term of energy.

We observe that for short pulse durations, the iterative DRAG-KAM algorithm \textcolor{black}{behaves as a superasymptotic series and} reaches a performance plateau after a few iterations. 
The number of iterations required to reach this plateau increases with the pulse duration. 

Concerning the experimental implementation of the proposed control pulse, we have made a spectral analysis with a typical transmon  (see the SM  \cite{SM}), since high-frequency components are expected to be degraded on a realistic hardware.
\textcolor{black}{We find specifically for the $\alpha$-RIO pulse that, for the shortest pulses reaching an infidelity of $-4$, the preservation of this threshold requires a cutoff of at least $250$~MHz for the hyper-Gaussian envelope and 146~MHz for the Gaussian envelope.}
This is a consequence of the semi-analytic generation of the pulse, the higher cutoff for the hyper-Gaussian envelope reflecting its more abrupt edges compared to the smoother Gaussian shape.
We have also checked the accuracy that can be expected conducting a calculation using the Lindblad dynamics with standard transmon coherence times $T_1=200\mu$s, $T_2=135\mu$s and $d/2\pi=-300$MHz \cite{Motzoi2024}, \textcolor{black}{ obtaining infidelities around ${-3.5}$ for both the $\alpha$-RIO and $\alpha\delta$-RIO pulses for durations large enough to reach ${-4}$ closed-system infidelity \cite{SM}.}

The RDK solutions serve as efficient initial guesses for further optimization using GRAPE, converging up to eightfold faster and reaching a nominal infidelity (i.e. for $\alpha=\delta=0$) two orders of magnitude lower than a RIO seed, as shown in the SM.

When leakage arises both below and above the qubit space, e.g. for the qubit $\{|1\rangle,|2\rangle\}$ as proposed in a more general setting where one considers a qutrit system $\{|0\rangle,|1\rangle,|2\rangle\}$ \cite{QuditMotzoi} or more generally a qudit for a $d$-level manifold \cite{Qudit}, it was suggested to address such multiple leakage by a recursive DRAG pulse consisting of successive DRAG applications featuring multi-derivative shaping \cite{Motzoi2024,QuditMotzoi,FeiYan}. 
Our technique can globally treat such multiple transition leakage at any order in a single step by an appropriate choice of the $W_j$'s (see the SM \cite{SM}), \textcolor{black}{while ensuring robustness against amplitude and detuning deviations}.
Although the framework is developed with anharmonic superconducting transmon qubits in mind, it can also be applied to other systems such as Bose Einstein condensates, where multiple leakage exists similarly \cite{JPB-Review,Guery-Odelin}. 

The present algorithm is versatile since any other protocol that has been developed in the qubit subspace can be integrated, \textcolor{black}{as long as it is sufficiently smooth in order to be differentiated. The time-reparametrization used to enforce $\Omega^{(1)}(t_i)=\Omega^{(1)}(t_f)=0$, independent of the original control boundaries, can introduce some limitations (see the End Matter for $\delta$-robustness example). Finally, time-reparametrization can be used to keep the perturbative parameter $\varepsilon$ sufficiently small.} One can mention among others universal robust control techniques which derive control pulses that are robust to any class of systematic unknown errors \cite{Kiely} or based on geometric frameworks \cite{KangyuanYi,Deng2025}.

\section*{Acknowledgments}
We acknowledge support from the EUR-EIPHI Graduate School (17-EURE-0002), the QuanTEdu-France project (ANR-22-CMAS-0001), the Conseil R\'egional de Bourgogne-Franche-Comt\'e, and Dijon M\'etropole. We acknowledge additional support from the TeLeMAQ LabCom (ANR-25-LCV2-0010-01).
We thank Dominique Sugny and Hans-Rudolf Jauslin for many fruitful exchanges during the preparation of this work.

\vfill

\eject

\newpage

\section*{End Matter}
\textit{Iterative DRAG-KAM algorithm.}
We renormalize the Hamiltonian $\hat H=H/|d|$ and apply the change of variable $s=t/T$ with $T$ the characteristic time of interaction, that leads to the Schr\"odinger equation $i\frac{\partial}{\partial s} \vert \phi \rangle = T|d| \hat H \vert \phi \rangle$. The perturbative parameters are of two kinds: static $\varepsilon =\max(\vert \Delta\vert, \vert\Omega\vert)  /\vert d\vert\ll 1$ and dynamical $\varepsilon_T=\frac{1}{T\vert d\vert}\ll 1$, leading to

\begin{equation}
	\label{factoriz}
	T|d| \hat H=\frac{1}{\varepsilon_T}\hat H= \frac{1}{\varepsilon_T}(\hat H_0+\varepsilon \hat V) =: \frac{1}{\varepsilon_T}\hat H_1
\end{equation}
with the dimensionless elements
\begin{equation}
	\hat H_0=\begin{pmatrix}0 & 0 &  0 \cr 
		0 & 0 & 0 \cr
		0 & 0  & \hat d
	\end{pmatrix},\quad  \hat V= \begin{pmatrix}0 &  \bar{\hat\Omega}/2  &  0 \cr 
		\hat\Omega/2 &\hat \Delta & \lambda \bar{\hat\Omega}/2 \cr
		0 & \lambda\hat\Omega/2  & 2\hat \Delta
	\end{pmatrix},
\end{equation}
where $\hat d = d/\vert d\vert$, $\hat \Omega = \Omega/\vert d\vert$ and  $\hat \Delta = \Delta /\vert d\vert$. 
We consider a situation with fast dynamics such that $\varepsilon$ is of order $\varepsilon_T$: $\varepsilon\sim\varepsilon_T$. 

At each iteration $N$, \textcolor{black}{we apply} $N$ KAM transformations, $\mathcal{T}_N = T_1T_2\dots T_N$, to the Hamiltonian. Each transformation is defined as $T_M = \exp(\varepsilon^M W_M)$, $M=1,2\dots N$. We consider also $N$ terms for the perturbation $\varepsilon \hat V = \sum_{M=1}^{N} \varepsilon^M \hat V^{(M)}$. At iteration $N$, all the parameters except $\hat V^{(N)}$ and $W_N$ are already known from previous iterations. Our goal is to choose $\hat V^{(N)}$ and $W_N$ such that the transformed Hamiltonian is of the form 
\begin{align}
	&\frac{1}{\varepsilon_T} \hat H_{N,t} = \mathcal{T}_N^\dagger\frac{ \hat H _N}{\varepsilon_T} \mathcal{T}_N - i \mathcal{T}_N^\dagger\dot{\mathcal{T}_N} \nonumber\\
	&=\frac{1}{\varepsilon_T} \begin{pmatrix}
		0 & \varepsilon \hat{\bar\Omega}^{(1)}/2 &  0 \cr 
		\varepsilon\hat\Omega^{(1)}/2  &\varepsilon\hat \Delta^{(1)} & 0 \cr
		0& 0 & \hat d + o(\varepsilon)
	\end{pmatrix} + o(\varepsilon^{N+1}/\varepsilon_T).
\end{align}
This comes down to require
\begin{align}
	&[{\hat H}_{N,t}]_{ij}^{(N)} = [{\hat H_0},W_N]_{ij} +[{\hat H}_{N-1,t}]^{(N)}_{ij}+ {\hat V}^{(N)}_{ij}=0, \nonumber \\
	&\text{for } i,j \in \{0,1,2\}, \quad (i,j) \neq (2,2),
\end{align} 
where the superscript of any operator $A$ is defined through the expansion:
$ A = A^{(0)}+\varepsilon A^{(1)} +\varepsilon^2 A^{(2)} +\cdots$
This corresponds to the definition of $\hat V^{(N)}$:
\begin{subequations}
	\begin{align} 
		\label{VN01}
		&\hat V^{(N)}_{ij} = -[\hat H_{N-1,t}]^{(N)}_{ij} \quad \nonumber \\ &\text{for}~~ (i,j) \in \{(0,0), (0,1), (1,0), (1,1)\}, \quad N \geq2,
	\end{align}
	which ensures that the controls within the qubit subspace remain unchanged in order to produce the desired gate,
	and the equation defining the $W_N$:
	\begin{align}
		\label{VN2}
		& [\hat H_0, W_N]_{ij}= -[\hat H_{N-1,t}]^{(N)}_{ij} - \hat V^{(N)}_{ij} \nonumber \\
		&\text{for} ~ (i,j) \in \{(0,2), (1,2), (2,0), (2,1)\}, \quad N \geq2,
	\end{align}
\end{subequations}
which eliminates all terms that couple the qubit subspace to the leakage subspace.
Any elements of the $W_N$ matrices not specified by the above equation may be regarded as free parameters, provided that the conditions $W_N^\dagger = -W_N$ and $\mathcal{T}_{N}(t_i)=\mathcal{T}_{N}(t_f)=\unit$
are satisfied. In our analysis, we can define a single free function $b(t) := - \varepsilon [W_1]_{01}\times \hat d = \varepsilon [W_1]_{10}\times \hat d$, which we further choose to be proportional to $\hat \Omega^{(1)}$: $b=-\kappa \hat \Omega^{(1)}/2$. This defines a single free parameter $\kappa$ in the DRAG-KAM procedure, which will be used for optimizing the reduction of the leakage effect and the fidelity.

The process can be implemented numerically by calculating any order of any product accordingly: 
$ (A B)^{(M)} = \sum_{i=0}^{M} A^{(i)}B^{(M-i)}$.
More details about the iterative algorithm can be found in the Supplemental Material (SM) \cite{SM}.

\begin{figure}[t!]
	\centering
	\includegraphics[width=0.4\textwidth]{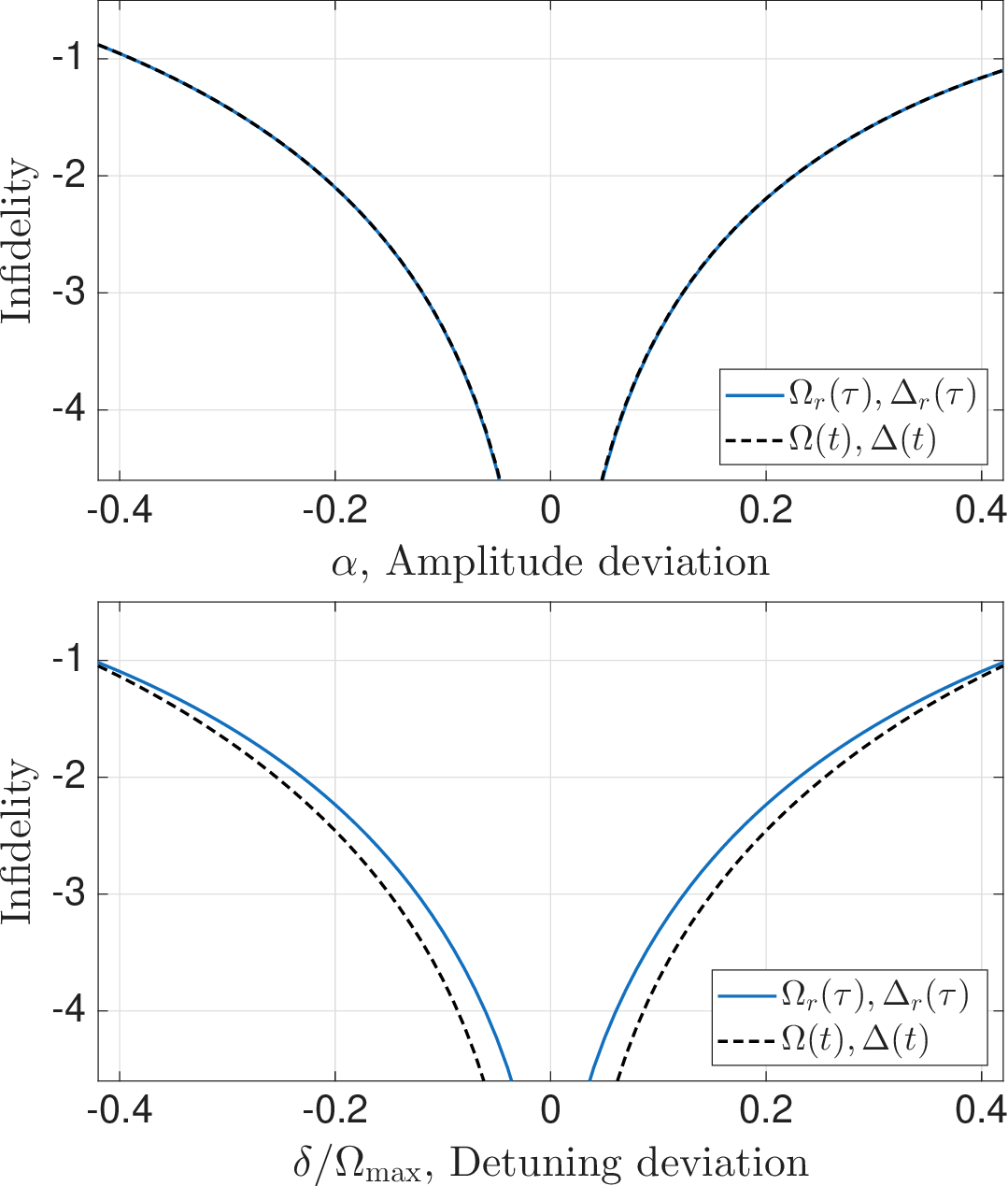}
	\caption{\textcolor{black}{Robustness profiles of the 2-level $\alpha\delta$-RIO pulse with duration $\sigma = 2.5\sigma_0$ comparing the reparametrized controls $\Omega_r(\tau),\Delta_r(\tau)$ to the original controls $\Omega(t),\Delta(t)$. Top: amplitude deviation $\alpha$. Bottom: detuning deviation $\delta/\Omega_{\max}$.}}
	\label{fig:Rob_timerepar}
\end{figure}

\textit{Time reparametrization.}
We define a new time variable $\tau$, such that $t = g(\tau)$, and reparametrize the control field $\Omega(t)$ using typically a  \textcolor{black}{Gaussian shape}:
$\Omega_r(\tau) = \Omega_{r0} \exp(-(\tau/\sigma)^{\textcolor{black}{2}} )$,
This smooth pulse shape guarantees vanishing boundaries at the beginning and end of the pulse, thereby eliminating discontinuities that would otherwise arise from sudden switching. To ensure the consistency of the dynamics, we require that the pulse area is preserved and that the initial and final quantum states remain unchanged. The latter is achieved by ensuring that the new time boundaries satisfy $\tau_i = g^{-1}(t_i)$ and $\tau_f = g^{-1}(t_f)$.
From the Schrödinger equation in the rescaled time frame, the new control fields are related to the original ones via:
\begin{equation}
	\Omega_r(\tau) = \textcolor{black}{\dot g}(\tau)\Omega(g(\tau)), \quad \Delta_r(\tau) = \textcolor{black}{\dot g}(\tau)\Delta(g(\tau)).
\end{equation}
Finally, from the area preservation condition, a relation between $\sigma$ and $\Omega_{r0}$ is derived as:
\begin{equation}\label{eq:timereparsig}
	\sigma = \frac{\textcolor{black}{2}}{2 \Omega_{r0} \Gamma(1/\textcolor{black}{2})} \int_{t_i}^{t_f} \Omega(t) \, dt.
\end{equation}
\textcolor{black}{ The time reparametrization is an exact change of variables and therefore leaves the nominal gate fidelity unchanged. However, it slightly degrades the $\delta$ robustness, while the $\alpha$ profile is unaffected, as shown in Fig.~\ref{fig:Rob_timerepar}, since $\Omega(t)\to(1+\alpha)\Omega(t)$ reparametrizes trivially into $\Omega_r(\tau)\to(1+\alpha)\Omega_r(\tau)$.}

\textcolor{black}{The loss in $\delta$ robustness originates from how the detuning deviation is applied. As in the plot, we apply the same normalized deviation $x\equiv\delta/\Omega_{\mathrm{max}}$ to both detuning profiles:
	\begin{subequations}
		\begin{align}
			\Delta(t) &\to \Delta(t) + x\max|\Omega|, \label{eq:dev_t}\\
			\Delta_r(\tau) &\to \Delta_r(\tau) + x\max|\Omega_r|. \label{eq:dev_tau}
		\end{align}
	\end{subequations}
	Reparametrizing equation \eqref{eq:dev_t} gives
	\begin{equation}
		\dot g(\tau)\big[\Delta(g(\tau)) + x\max|\Omega|\big] = \Delta_r(\tau) + x\,\dot g(\tau)\max|\Omega|,
	\end{equation}
	which in general differs from equation \eqref{eq:dev_tau}, since in general
	\begin{equation}
		\dot g(\tau)\max_\tau|\Omega(g(\tau))| \;\neq\; \max_\tau|\Omega_r(\tau)| = \max_\tau|\dot g(\tau)\Omega(g(\tau))|.
	\end{equation}
	The two normalized sweeps therefore correspond to different absolute detuning deviations in the two frames, which is the origin of the slightly different $\delta$ robustness profiles observed in Fig.~\ref{fig:Rob_timerepar}.}

\begin{figure}[h]
	\centering
	\includegraphics[width=0.9\linewidth]{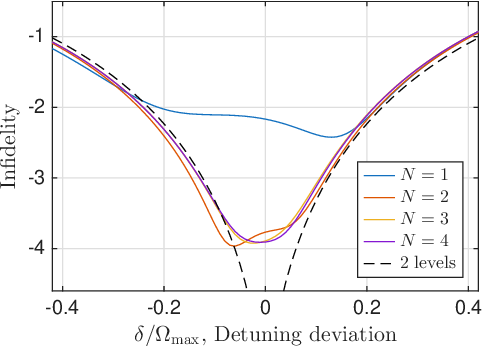}
	\caption{Robustness profile versus the detuning at each iteration up to 4 iterations for $\alpha\delta$-RIO with $\sigma = 2.5 \sigma_0$ for which the infidelity ${-4}$ is approximately reached.}\label{fig:Robustness_alphadeltario_delta}
\end{figure}

\textit{Robustness profile in $\delta$.}
In the main text, the $\alpha$ profile is analyzed. In this section, we show the $\delta$ profile in Fig. \ref{fig:Robustness_alphadeltario_delta} using the same notations as in the main text. This proves that here again DRAG-KAM recovers the robustness of two-level RIO controls after three iterations ($N=3$). 

\clearpage
\onecolumngrid
\begin{center}
{\large\bfseries SUPPLEMENTARY MATERIAL}
\end{center}
\twocolumngrid
\setcounter{section}{0}\setcounter{equation}{0}\setcounter{figure}{0}\setcounter{table}{0}
\renewcommand{\thesection}{\Roman{section}}
\renewcommand{\theequation}{\arabic{equation}}
\renewcommand{\thefigure}{\arabic{figure}}
\renewcommand{\thetable}{\arabic{table}}

\section{Structure of the supplementary material}
We first give in Section II the details of the step-by-step derivation of the iterative DRAG-KAM algorithm for a three-level system. 
In Section III, we extend the framework to the $|1\rangle$--$|2\rangle$ qubit encoding, demonstrating that the algorithm simultaneously suppresses leakage to both adjacent levels in a single global framework. Section IV is devoted to the discussion of the effect of pulse shape on spectral properties, comparing hyper-Gaussian and Gaussian envelopes in terms of bandwidth requirements and robustness. In Section V, we analyze the GRadient Ascent Pulse Engineering (GRAPE) method applied on RIO or RDK pulses, showing that the RDK pulse as an initial condition increases the performance of GRAPE in term of nominal fidelity (i.e. for $\alpha=\delta=0$). Section VI provides the dynamics including the decoherence with standard transmon coherence times by a Lindblad master equation.

\section{Iterative DRAG-KAM algorithm: Step by step explanation for $3$-level system}
\subsection{First iteration}
At the first iteration we consider the control Hamiltonian as the first order perturbation:  $\varepsilon V^{(1)}$. The Hamiltonian writes as
\begin{equation}
	\frac{1}{\varepsilon_T}\hat H_1 = \frac{1}{\varepsilon_T}( \hat H_0 + \varepsilon \hat V^{(1)}),
\end{equation}
where
\begin{align}
	\hat H_0=\begin{pmatrix}0 & 0 &  0 \cr 
		0 & 0 & 0 \cr
		0 & 0  & \hat d
	\end{pmatrix}
\end{align}
and
\begin{align}
	\hat V^{(1)}= \begin{pmatrix}0 &  \hat{\bar\Omega}^{(1)}/2  &  0 \cr 
		\hat\Omega^{(1)}/2 &\hat \Delta^{(1)} & \lambda \hat{\bar\Omega}^{(1)}/2 \cr
		0 & \lambda\hat\Omega^{(1)}/2  & 2\hat \Delta^{(1)}
	\end{pmatrix}.
\end{align}
The controls $\hat \Omega^{(1)} = \Omega^{(1)}/ \vert d \vert, \hat \Delta^{(1)}= \Delta^{(1)}/ \vert d \vert$ and the anharmonicity $\hat d = d/ \vert d \vert$ are rescaled through the  normalization with $\vert d \vert$. We omit the hats in this supplementary material for ease of notation, but keep them in the main text. Without loss of generality $\Omega^{(1)}$ is considered real. The-off diagonal terms $\lambda\Omega^{(1)}/2$ couple the states 0-2 and 1-2 causing leakage to the upper state. 
From the cohomology equations
\begin{subequations}
	\begin{align}
		D_{1} &=[H_{0},W_{1}]+V_{1},\\
		[H_{0},D_{1}] &=0,
	\end{align}
\end{subequations}
and their solutions
\begin{subequations}
	\label{sol_coh}
	\begin{eqnarray}
		D_{1} &=&\sum_{\bnu,j,j^{\prime }}|\bnu,j\rangle \langle \bnu%
		,j|V_{1}|\bnu,j^{\prime }\rangle \langle \bnu,j^{\prime }|,  \label{solWD} \\
		W_{1} &=&-\sum_{\bnu,j,j^{\prime },\bnu^{\prime }\not=\bnu}{\frac{|\bnu%
				,j\rangle \langle \bnu,j|V_1|\bnu^{\prime },j^{\prime }\rangle \langle \bnu%
				^{\prime },j^{\prime }|}{\lambda _{\bnu}^{0}-\lambda
				_{\bnu^{\prime }}^{0}}},
		\label{solWD_}
	\end{eqnarray}
\end{subequations}
the transformation $T_1 = \exp(\varepsilon W_1)$ takes the form
\begin{align}
	W_1= \frac{1}{d} \begin{pmatrix}0 & -b & 0 \cr 
		b & 0 & -\lambda a \cr
		0 & \lambda a & 0
	\end{pmatrix},\quad
	a=-\Omega^{(1)}/2, 
\end{align}
where we have added a free parameter $b$ inside the qubit block, which commutes with $H_0$.
This transformation cancels out the off-diagonal terms of order $\varepsilon$  inside the transformed Hamiltonian, increasing the order of leakage to $\varepsilon^2$, such that the transformed Hamiltonian is
\begin{align}
	\frac{1}{\varepsilon_T} H_{1,t} = \frac{1}{\varepsilon_T}\begin{pmatrix}0 &\varepsilon\Omega^{(1)}/2 &  0 \cr 
		\varepsilon\Omega^{(1)}/2  & \varepsilon\Delta^{(1)} & 0 \cr
		0& 0  &d+o(\varepsilon)
	\end{pmatrix}  + o(\varepsilon^2/\varepsilon_T).
\end{align}
We choose for convenience $b$ proportional to $a$ and $\kappa:=b/a$ as a free parameter.
To ensure that the initial and final states remain the same in the transformed frame, the initial and final transformations must be identity: 
\begin{align}
	T_1(t_i) = e^{\varepsilon W_1(t_i)}=T_1(t_f) =e^{\varepsilon W_1(t_f)}=\id,
\end{align} 
or equivalently 
\begin{align}
	a(t_i)=a(t_f)=0 \implies \Omega^{(1)}(t_i) = \Omega^{(1)}(t_f) = 0.
\end{align}
At this stage, we have not yet introduced any corrective controls, therefore the final fidelity is not improved, thus iteration 1 gives the same results as the three-level system does without any leakage suppression.

\subsection{Second iteration}
In order to decrease the leakage and to preserve the effective form of the original coupling within the qubit subspace, we add a corrective control term $\varepsilon^2 V^{(2)}$ to the Hamiltonian:
\begin{equation}
	\frac{1}{\varepsilon_T} H_2 = \frac{1}{\varepsilon_T}(H_0 + \varepsilon V^{(1)}+\varepsilon^2 V^{(2)}),
\end{equation}
where
\begin{align}
	&V^{(2)} = \begin{pmatrix}0 & \bar\Omega^{(2)}/2 &  0 \cr 
		\Omega^{(2)}/2  & \Delta^{(2)} & \lambda\bar\Omega^{(2)}/2 \cr
		0& \lambda\Omega^{(2)}/2  & 2\Delta^{(2)}
	\end{pmatrix},
\end{align}
and transform it twice with $\mathcal{T}_2 = T_1 T_2$, where $T_2 = \exp(\varepsilon^2 W_2)$ and
\begin{align}
	W_2= \begin{pmatrix}0 & 0 & \bar\mu_2 \cr 
		0 & 0 & \bar\xi_2 \cr
		-\mu_2 & -\xi_2 & 0
	\end{pmatrix}.
\end{align}
The shape of $W_2$ is constructed according to the cohomology solution (\ref{solWD_}).
The transformed Hamiltonian is
\begin{align}
	\frac{1}{\varepsilon_T} H_{2,t} = \mathcal{T}_2^\dagger\frac{ H_2}{\varepsilon_T} \mathcal{T}_2 - i \mathcal{T}_2^\dagger\dot{\mathcal{T}_2}.
\end{align}
We choose $V^{(2)}$ and $W_2$ such that the transformed Hamiltonian is of the form
\begin{align}
	\frac{1}{\varepsilon_T} H_{2,t} = \frac{1}{\varepsilon_T}\begin{pmatrix}0 &\varepsilon\Omega^{(1)}/2 &  0 \cr 
		\varepsilon\Omega^{(1)}/2  & \varepsilon\Delta^{(1)} & 0 \cr
		0& 0  &d+o(\varepsilon)
	\end{pmatrix} + o(\varepsilon^3/\varepsilon_T).
\end{align}
Any unwanted terms causing leakage or modifying the controls of the qubit block are now of order $o(\varepsilon^3/\varepsilon_T)$. Equivalently, this condition can be expressed as
\begin{align}
	[H_{2,t}]^{(2)}_{ij} =[\mathcal{T}_2^\dagger H_2 \mathcal{T}_2 - i \varepsilon_T\mathcal{T}_2^\dagger\dot{\mathcal{T}_2}]^{(2)}_{ij}= 0,\nonumber \\
	\text{for}\quad i,j \in \{0,1,2\}, \quad (i,j) \neq (2,2) .
\end{align}
Thanks to the coefficient $\varepsilon_T$, the $T_2$ transformation does not introduce any terms in the dynamical term, which can be simplified as:
\begin{align}
	[\varepsilon_T\mathcal{T}_2^\dagger\dot{\mathcal{T}_2}]^{(2)}=[\mathcal{T}_2^\dagger\dot{\mathcal{T}_2}]^{(1)} = [T_1^\dagger\dot T_1]^{(1)} =[\varepsilon_T T_1^\dagger\dot T_1]^{(2)}.
\end{align} 
The other term can be simplified as well:
\begin{align}
	[\mathcal{T}_2^\dagger H_2 \mathcal{T}_2]^{(2)}=[\mathcal{T}_2^\dagger (H_0 + \varepsilon V^{(1)}) \mathcal{T}_2]^{(2)} + [\mathcal{T}_2^\dagger \varepsilon^2 V^{(2)} \mathcal{T}_2]^{(2)} \nonumber\\
	=[\mathcal{T}_2^\dagger H_1 \mathcal{T}_2]^{(2)} + V^{(2)} = [T_1^\dagger H_1 T_1]^{(2)} + [H_0,W_2]+ V^{(2)}.
\end{align}
Finally adding both simplifications together we arrive at
\begin{align}
	[H_{2,t}]^{(2)} = [H_0,W_2] + V^{(2)} + [H_{1,t}]^{(2)}.
\end{align}
The unknowns to be determined are $V^{(2)}$ and $W_2$, where the latter only appears through its commutator with $H_0$.
The process is implemented in two steps: i) Cancel any elements in the 0-1 block that are of  $o(\varepsilon^2/\varepsilon_T)$ order, giving $V^{(2)}$; ii) Cancel the off-diagonal terms of  $o(\varepsilon^2/\varepsilon_T)$ order, giving $W_2$:

i) Due to the structure of $W_2$ and $H_0$ the commutator $[H_0,W_2]$ is 0 inside the qubit block, therefore from $[H_{2,t}]^{(2)}=0$ inside the qubit block we get a simple solution to $V^{(2)}$:  
\begin{align}
	V^{(2)}_{ij} = -[H_{1,t}]^{(2)}_{ij}, \text{  for}\quad i,j \in \{0,1\}.
\end{align}
This gives $\Omega^{(2)}$ and $\Delta^{(2)}$, which construct the full correction $V^{(2)}$.

ii) The transformation $W_2$ is found from canceling the coupling terms:
\begin{align}
	\label{W_2_solution}
	&[H_0, W_2]_{ij} = -[H_{1,t}]^{(2)}_{ij} - V^{(2)}_{ij},\nonumber\\ &\text{for  } (i,j)\in\{(0,2),(1,2),(2,0),(2,1)\}. 
\end{align}
As $H_0$ is already known, equation (\ref{W_2_solution}) is enough to determine $\mu_2$ and $\xi_2$ and in turn $W_2$.
\subsection{N-th iteration}

At the $N$th iteration, we apply $\mathcal{T}_N = T_1 T_2 \dots T_N$ transformations, where $T_M = \exp(\varepsilon^M W_M)$ and add  $N$ corrections: $V = \sum_{M=1}^N \varepsilon^M V^{(M)}$. The transformed Hamiltonian
\begin{align}
	\frac{1}{\varepsilon_T} H_{N,t} = \frac{1}{\varepsilon_T} \begin{pmatrix}0 &\varepsilon\Omega^{(1)}/2 &  0 \cr 
		\varepsilon\Omega^{(1)}/2  & \varepsilon\Delta^{(1)} & 0 \cr
		0& 0  &d+o(\varepsilon)
	\end{pmatrix} + o\Biggl(\frac{\varepsilon^{N+1}}{\varepsilon_T}\Biggr)
\end{align}
is block-diagonal up to order $o(\varepsilon^{N+1}/\varepsilon_T)$. Everything besides $W_N$ and $V^{(N)}$ are already known from previous iterations. In fact the new transformation $T_N$ and the new correction $V^{(N)}$ only modify terms of order $o(\varepsilon^{N}/\varepsilon_T)$ 
and higher, therefore we only require
\begin{align}
	[H_{N,t}]^{(N)}_{ij} =[\mathcal{T}_N^\dagger H_N \mathcal{T}_N - i \varepsilon_T\mathcal{T}_N^\dagger\dot{\mathcal{T}_N}]^{(N)}_{ij}= 0,\nonumber\\
	\text{for}\quad i,j \in \{0,1,2\}, \quad (i,j) \neq (2,2) .
\end{align}
By making the same simplifications as we did for the second iteration, this gives the relations defining $V^{(N)}$ and $W_N$:
\begin{align} 
	&V^{(N)}_{ij} = -[H_{N-1,t}]^{(N)}_{ij} \quad \nonumber \\ &\text{for}~~ (i,j) \in \{(0,0), (0,1), (1,0), (1,1)\}, \quad N \geq2,
\end{align}
and
\begin{align}
	\label{eq:WNsolution}
	& [H_0, W_N]_{ij}= -[H_{N-1,t}]^{(N)}_{ij} - V^{(N)}_{ij} \nonumber \\
	&\text{for} ~ (i,j) \in \{(0,2), (1,2), (2,0), (2,1)\}, \quad N \geq2, 
\end{align}
where the corrective controls are 
\begin{align}
	& V^{(N)} = \begin{pmatrix}0 & \bar\Omega^{(N)}/2 &  0 \cr 
		\Omega^{(N)}/2  & \Delta^{(N)} & \lambda\bar\Omega^{(N)}/2 \cr
		0& \lambda\Omega^{(N)}/2  & 2 \Delta^{(N)} 
	\end{pmatrix},
\end{align}
and $W_N$ is  of the form
\begin{align}
	W_N=\begin{pmatrix}0 & 0 & \bar\mu_N \cr 
		0 & 0 & \bar\xi_N \cr
		-\mu_N & -\xi_N & 0
	\end{pmatrix}.
\end{align}

\section{DRAG-KAM for the $|1\rangle$--$|2\rangle$ qubit}

We here apply the iterative DRAG-KAM algorithm to a weakly anharmonic oscillator in which the states $|1\rangle$ and $|2\rangle$ serve as the qubit. In this system, leakage exists both below or above the qubit with a detuning of opposite sign. This strategy can be viewed in a more general setting where one considers a qutrit system $\{|0\rangle,|1\rangle,|2\rangle\}$ \cite{QuditMotzoi}, or more generally a qudit for a $d$-level manifold \cite{Qudit}.
It was suggested in \cite{Motzoi2024,QuditMotzoi} to address a multiple leakage by a recursive multistep DRAG pulse. Here we show that our technique can globally treat it at any order in a single step by an appropriate choice of the $W_j$'s. 

\begin{figure}[h]
	\centering
	\begin{minipage}[b]{0.4\textwidth}
		\centering
		\includegraphics[width=\textwidth]{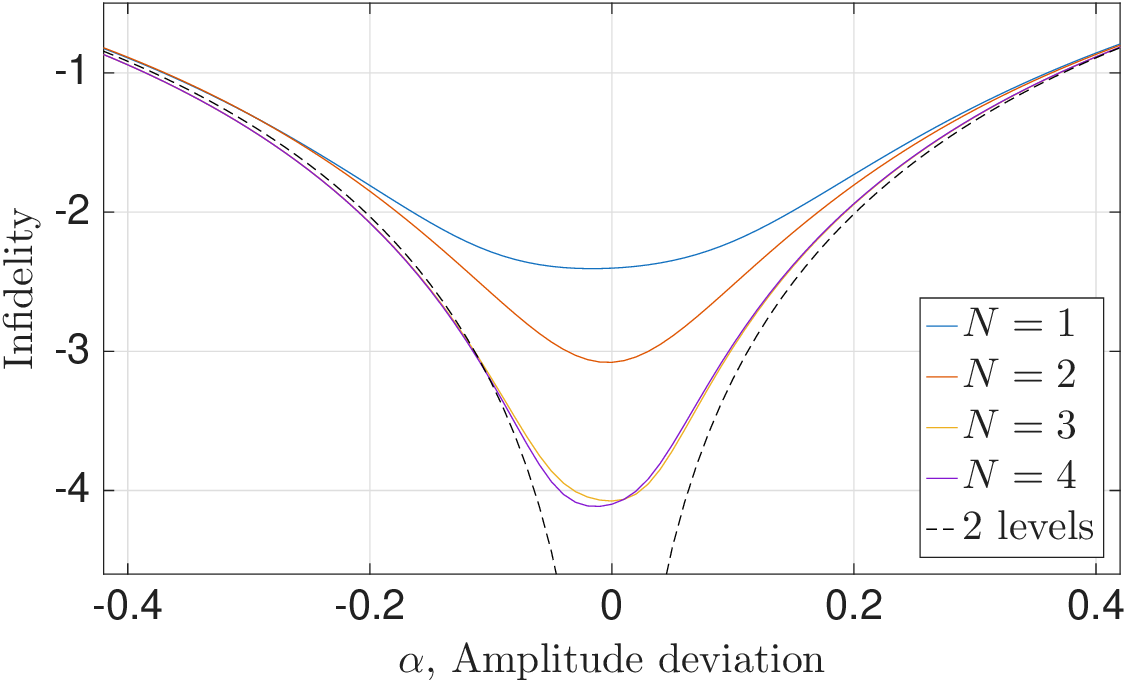}
	\end{minipage}
	\hfill
	\begin{minipage}[b]{0.4\textwidth}
		\centering
		\includegraphics[width=\textwidth]{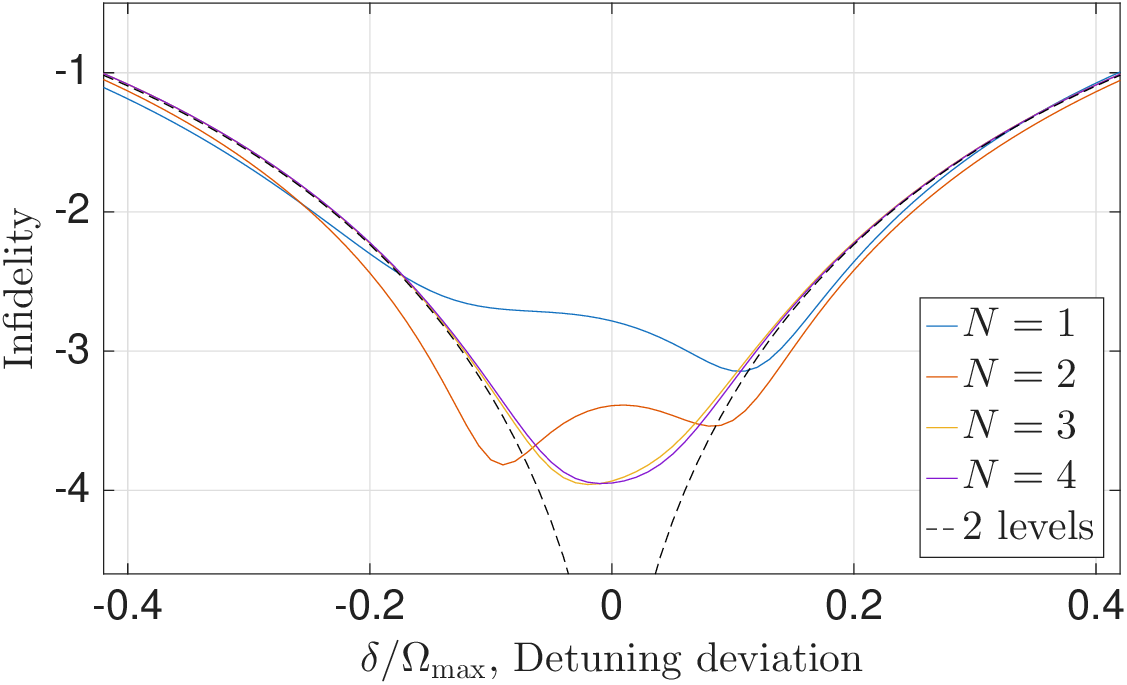}
	\end{minipage}
	\caption{Robustness profiles for $N=4$ iterations for the $|1\rangle$--$|2\rangle$ qubit. 
		Top: amplitude deviation profile for the $\alpha$-RIO pulse at $\sigma = 2.2\,\sigma_0$. 
		Bottom: detuning deviation profile for the $\alpha\delta$-RIO pulse at $\sigma = 2.8\,\sigma_0$, with $\Omega_{\max} = \max (\lambda_2\Omega^{(1)})$.}
	\label{fig:DRAG_robustness12}
\end{figure}

We simultaneously decouple them from both the lower level $|0\rangle$ and the higher level $|3\rangle$ in a single global procedure. The same framework can be applied to any pair of adjacent levels $|k\rangle$--$|k+1\rangle$ in the anharmonic ladder, providing a systematic way to isolate a pair.
The 4-level Hamiltonian $H_1 = H_0 + \varepsilon V^{(1)}$ then writes
\begin{equation}
	H_0 = \mathrm{diag}(d,\, 0,\, 0,\, d),
\end{equation}
\begin{equation}
	V^{(1)} = \frac{1}{2}\begin{pmatrix}
		0 & \bar\Omega^{(1)} & 0 & 0 \\[4pt]
		\Omega^{(1)} & 2\tilde\Delta^{(1)} & \lambda_2\bar\Omega^{(1)} & 0 \\[4pt]
		0 & \lambda_2\Omega^{(1)} & 4\tilde\Delta^{(1)} & \lambda_3\bar\Omega^{(1)} \\[4pt]
		0 & 0 & \lambda_3\Omega^{(1)} & 6\tilde\Delta^{(1)}
	\end{pmatrix},
\end{equation}
where $\tilde{\Delta}^{(1)}$ is the detuning between the driving frequency $\omega$ and the  $|1\rangle\to|2\rangle$ transition frequency: $\tilde{\Delta}^{(1)} = \omega_2 - \omega_1 - \omega = \Delta^{(1)} + d$, where $\omega_1$ is the frequency of $|1\rangle$ and $\omega_2 = 2\omega_1 + d$ that of $|2\rangle$. The parameters $\lambda_j = \sqrt{j}$ and the anharmonicity of level 3 is $d_3 = 3d = \omega_3 - 3\omega_1$. The first contact transformation $W_1$ according to the KAM solution~\eqref{solWD_} writes:
\begin{equation}
	W_1 = \frac{1}{2d_2}\begin{pmatrix}
		0 & -\bar\Omega^{(1)} & 0 & 0 \\
		\Omega^{(1)} & 0 & 0 & 0 \\
		0 & 0 & 0 & \lambda_3\bar\Omega^{(1)} \\
		0 & 0 & -\lambda_3\Omega^{(1)} & 0
	\end{pmatrix},
\end{equation}
After $N$ iterations, we require the effective Hamiltonian to be block-diagonal
\begin{equation}
	H_{N,t} = \begin{pmatrix}
		d + o(\varepsilon) & 0 & 0 & 0 \\
		0 & \varepsilon\tilde\Delta^{(1)} & \frac{\lambda_2\varepsilon\bar\Omega^{(1)}}{2} & 0 \\
		0 & \frac{\lambda_2\varepsilon\Omega^{(1)}}{2} & 2\varepsilon\tilde\Delta^{(1)} & 0 \\
		0 & 0 & 0 & d + o(\varepsilon)
	\end{pmatrix}+o(\varepsilon^{N+1}).
\end{equation}
The control corrections then write as
\begin{subequations}
	\begin{align}
		\lambda_2\Omega^{(N)}/2 = [H_{N-1,t}]^{(N)}_{21}, \\
		\tilde\Delta^{(N)} = [H_{N-1,t}]^{(N)}_{22} - [H_{N-1,t}]^{(N)}_{11}.
	\end{align}
\end{subequations}
The contact transformations $W_{N\geq 2}$ are anti-Hermitian matrices of the form
\begin{equation}
	W_{N\geq 2} = \begin{pmatrix}
		0 & W_{01} & W_{02} & 0 \\
		-\bar W_{01} & 0 & 0 & W_{12} \\
		-\bar W_{02} & 0 & 0 & W_{13} \\
		0  & -\bar W_{12} & -\bar W_{13} & 0
	\end{pmatrix},
\end{equation}
whose components are determined by Eq.~\eqref{eq:WNsolution} for $(i,j)\in{(0,1), (0,2), (1,2), (1,3)}$.

An infidelity of ${-4}$ is attained for $\alpha$-RIO and $\alpha\delta$-RIO at durations $\sigma = 2.2\,\sigma_0$ and $\sigma = 2.8\,\sigma_0$, respectively, only slightly longer than the $|0\rangle$--$|1\rangle$ case ($2\,\sigma_0$ and $2.5\,\sigma_0$). The robustness profiles in Fig.~\ref{fig:DRAG_robustness12} recover the 2-level robustness in both amplitude and detuning deviations, demonstrating that simultaneous suppression of leakage to both adjacent levels comes at only a modest cost in pulse duration.

\section{The effect of pulse shape on the spectral properties}

\begin{figure}[h]
	\centering
	\begin{minipage}[b]{0.48\textwidth}
		\centering
		\includegraphics[width=\textwidth]{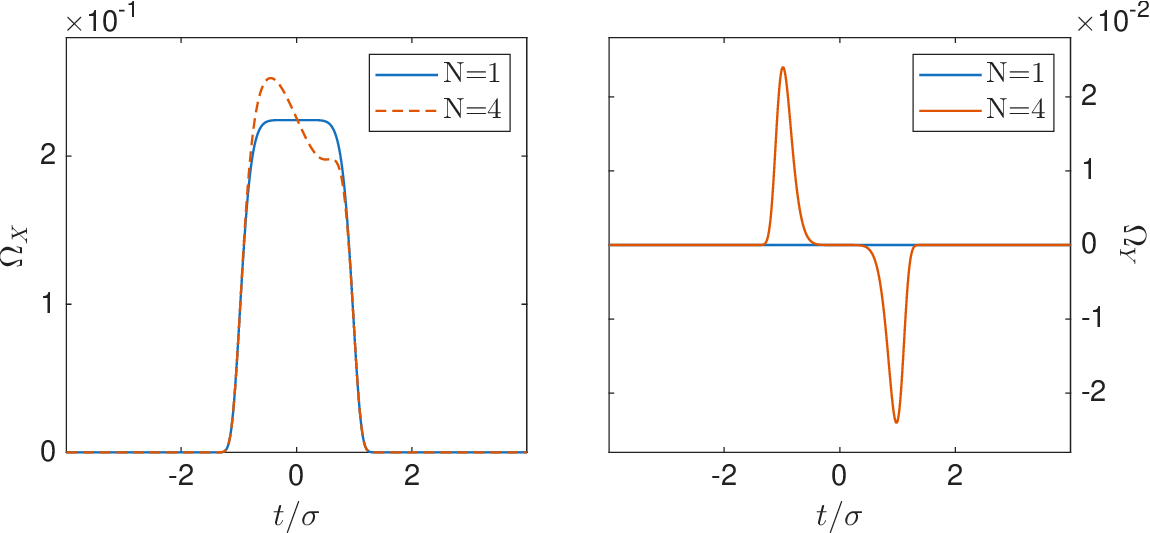}
		\caption{Temporal profile of $\Omega(t)= \Omega_X(t) + i \Omega_Y(t)$ for the hyper-Gaussian ($n=8$) $\alpha$-RIO pulse with $\sigma = 2.2\sigma_0$.}
		\label{fig:Omegatimehyper}
	\end{minipage}
	\hfill
	\begin{minipage}[b]{0.48\textwidth}
		\centering
		\includegraphics[width=\textwidth]{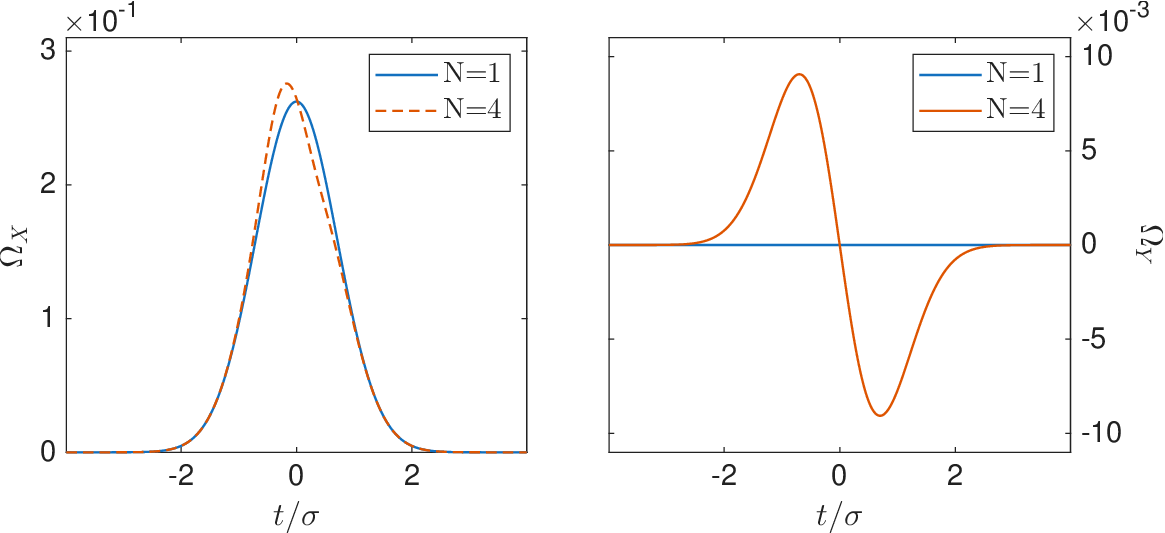}
		\caption{Temporal profile of $\Omega(t)= \Omega_X(t) + i \Omega_Y(t)$ for the Gaussian ($n=2$) $\alpha$-RIO pulse with $\sigma = 2\sigma_0$.}
		\label{fig:Omegatime}
	\end{minipage}
\end{figure}

\begin{figure}[h]
	\centering
	\begin{minipage}[b]{0.48\textwidth}
		\centering
		\includegraphics[width=\textwidth]{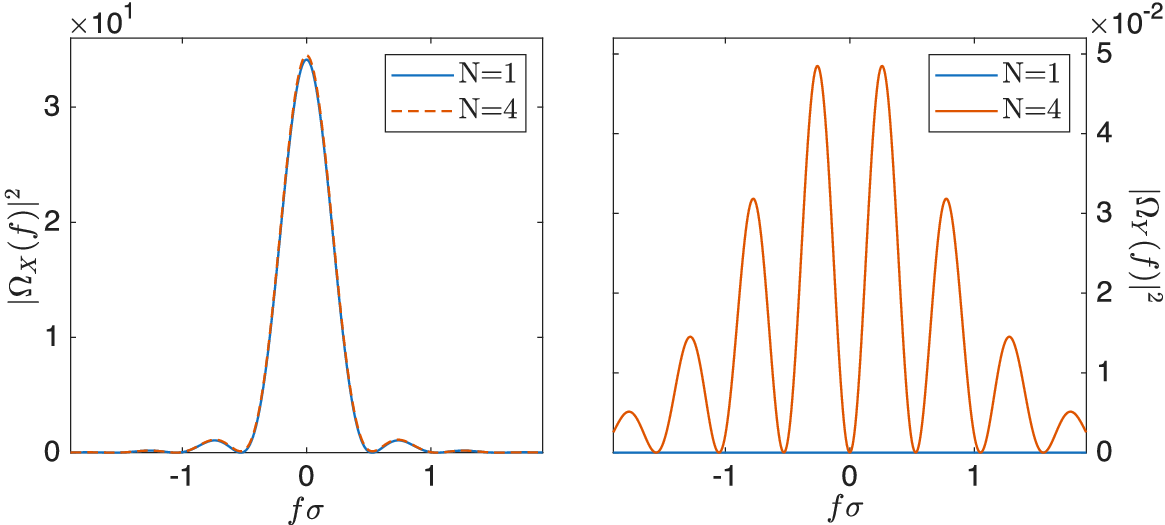}
		\caption{Spectral profile of $\Omega(t)$ for the hyper-Gaussian ($n=8$) $\alpha$-RIO pulse with $\sigma = 2.2\sigma_0$.}
		\label{fig:Omegafreqhyper}
	\end{minipage}
	
	\begin{minipage}[b]{0.48\textwidth}
		\centering
		\includegraphics[width=\textwidth]{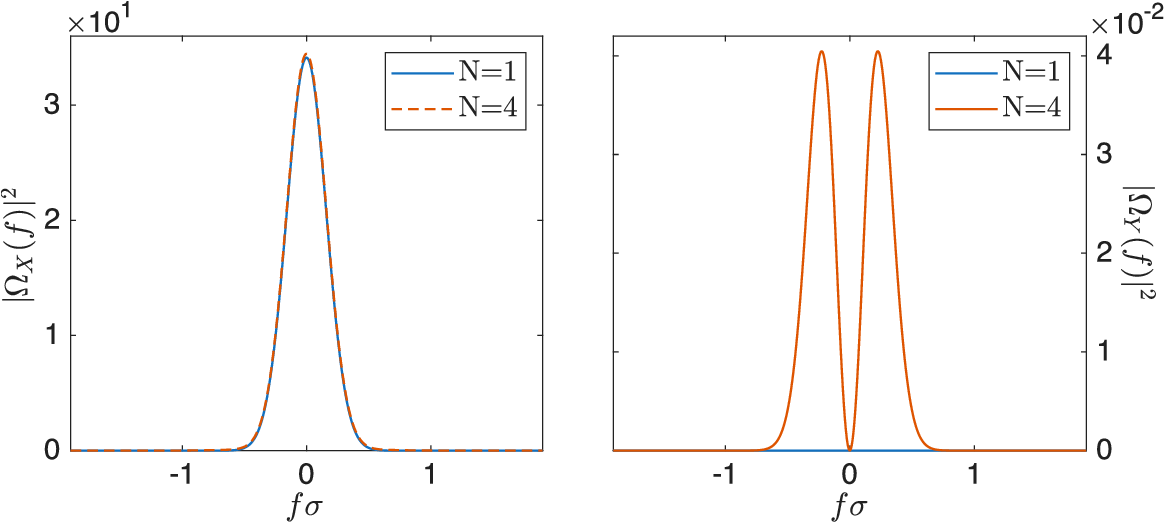}
		\caption{Spectral profile of $\Omega(t)$ for the Gaussian ($n=2$) $\alpha$-RIO pulse with $\sigma = 2\sigma_0$.}
		\label{fig:Omegafreq}
	\end{minipage}
\end{figure}

To mimic the original $\alpha$-RIO control \cite{Dridi}, which features a constant $\Omega(t)$, we use a pulse of hyper-Gaussian shape: $\Omega(t) = \Omega_0 \exp(-(t/\sigma)^n)$ \cite{ZhuJPB}. For $\sigma=2.2\sigma_0$ and $n=8$, the infidelity after applying the iterative DRAG-KAM algorithm improves from ${-2}$ to ${-4}$ at iteration $N=4$. We note that in our analysis we always consider the iteration $N=4$, which gives almost identical results as $N=3$. The modifications to the Rabi frequency are approximately $10\%$, as seen in Fig.~\ref{fig:Omegatimehyper}, while the detuning modifications are negligible. However, the abrupt edges of $\Omega(t)$ produce a spectral profile with significant high-frequency content, as shown in Fig.~\ref{fig:Omegafreqhyper}. This limitation can be avoided by choosing a Gaussian pulse with $n=2$. For $\sigma=2\sigma_0$, the same improvement in infidelity from approximately ${-2}$ to ${-4}$ is achieved at iteration $N=4$, with modifications of only $3\%$, as shown in Fig.~\ref{fig:Omegatime}. The spectral profiles of Fig.~\ref{fig:Omegafreq} are considerably narrower, free of high-frequency components. The trade-off is the effective pulse duration: despite its slightly smaller $\sigma$, the Gaussian envelope forces a longer effective duration, with $99\%$ of the total pulse energy contained within $5.1\,\sigma_0$ compared to $4.4\,\sigma_0$ for the hyper-Gaussian. 

\begin{table}[h]
	\centering
	\setlength{\tabcolsep}{7pt}
	\begin{tabular}{ccccccc}
		\hline
		$\sigma$ & $n$ & $T_{99\%} $&  $f_c$ (${-4}$) & $f_c$ (${-3}$) \\
		\hline
		
		$2.2\sigma_0$ & 8 & 4.4$\sigma_0$ &  250 MHz & 105 MHz \\
		
		$2\sigma_0$ & 2 & 5.1$\sigma_0$ &  146 MHz & 120 MHz \\
		
		\hline
	\end{tabular}
	\caption{Spectral bandwidth requirements for $\alpha$-RIO pulses at anharmonicity $d/2\pi = -300$ MHz, comparing hyper-Gaussian ($n=8$) and Gaussian ($n=2$) envelopes after $N=4$ iterations. All pulses are optimized such that the infidelity is slightly below ${-4}$. $T_{99\%}$ denotes the duration containing $99\%$ of the total pulse energy, and $f_c$ is the cutoff frequency below which the infidelity exceeds the given threshold (in parenthesis).}
	\label{tab:bandwidth}
\end{table}

\begin{figure}
	\centering
	\includegraphics[width=0.4\textwidth]{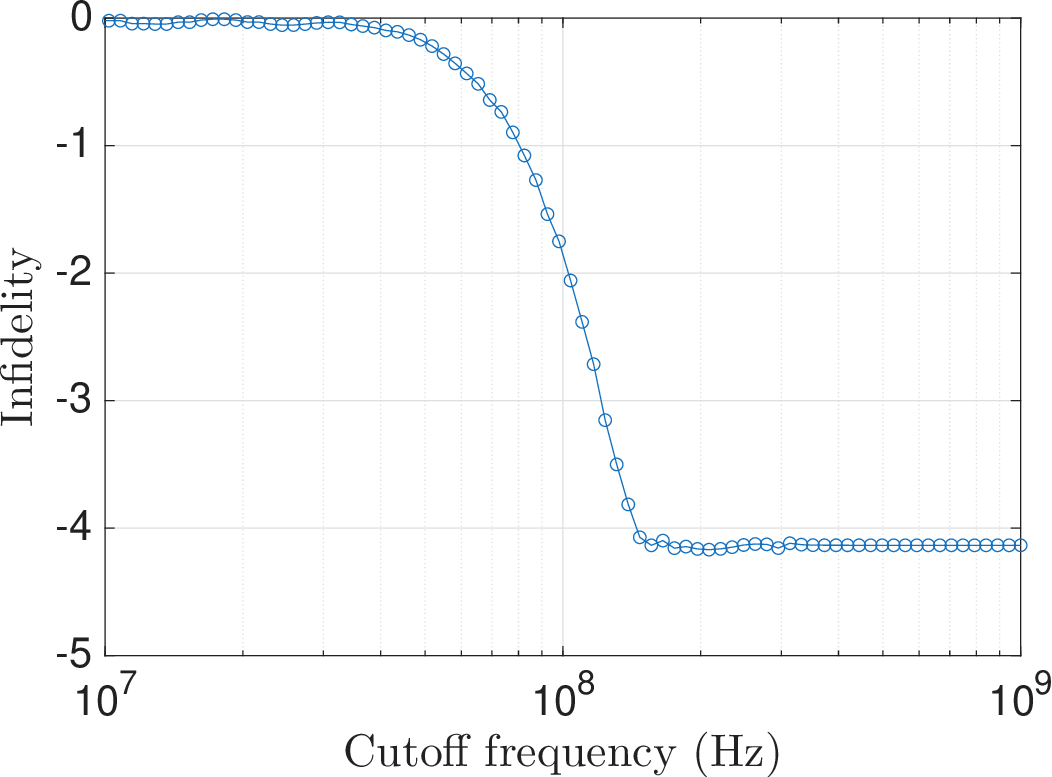}
	\caption{ Infidelity as a function of the low-pass filter cutoff frequency applied to $\Omega(t)$, for the Gaussian $\alpha$-RIO pulse after $N=4$ iterations with $\sigma=2\sigma_0\equiv 6.7$ns and $d/2\pi=-300$~MHz.}
	\label{fig:cutoff}
\end{figure}

\begin{figure}
	\centering
	\includegraphics[width=0.4\textwidth]{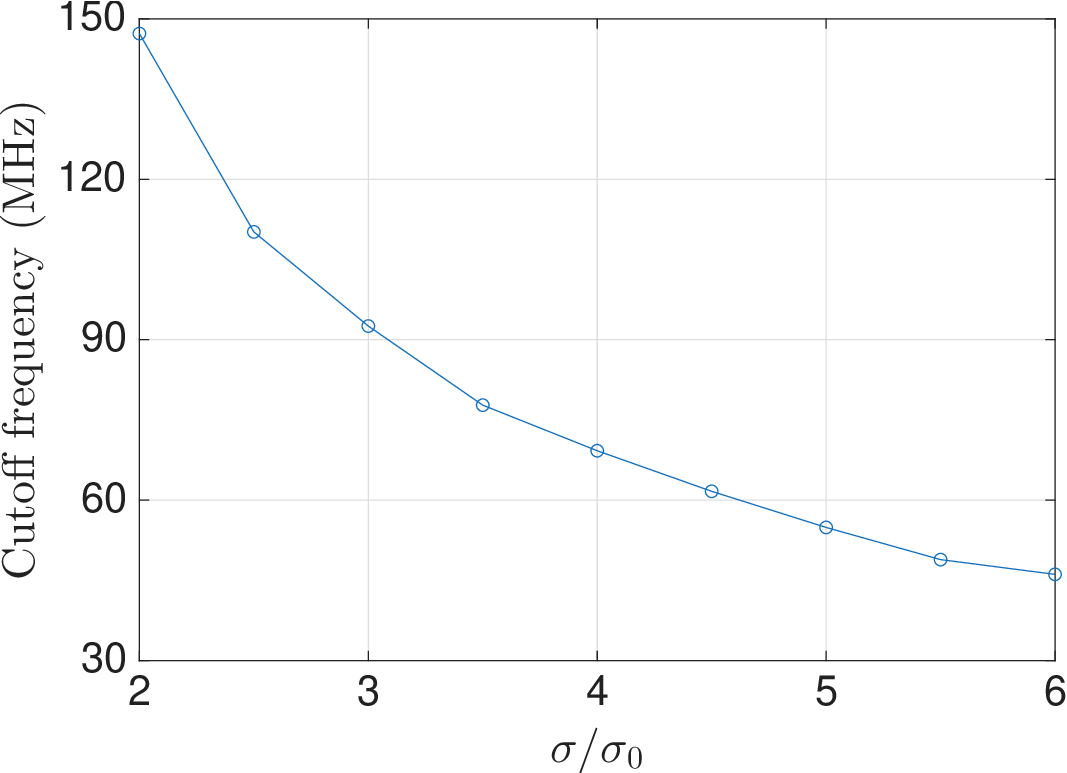}
	\caption{Cutoff frequency required to maintain infidelity below ${-4}$ at iteration $N=4$ as a function of the pulse duration $\sigma/\sigma_0$, for $\alpha$-RIO with a Gaussian envelope. The range shown corresponds to $\sigma = 6.7$--$20$~ns for $d/2\pi = -300$~MHz.}
	\label{fig:cutoff_vs_sigma}
\end{figure}

For an anharmonicity of $d/2\pi = -300$ MHz, the hyper-Gaussian ($n=8$) $\alpha$-RIO pulse with $\sigma = 2.2\,\sigma_0$ achieves an infidelity of ${-4.14}$, remains below ${-4}$ up to a bandwidth of 250 MHz, and degrades to ${-3}$ at 105 MHz. The Gaussian $\alpha$-RIO pulse with $\sigma = 2\,\sigma_0$ achieves a comparable infidelity of ${-4.13}$, yet retains ${-4}$ only up to 146 MHz and reaches ${-3}$ at 120 MHz. Despite requiring only $16\%$ longer duration, the Gaussian pulse has significantly reduced high-frequency content, making it better suited for bandwidth-limited experimental setups.
The cutoff frequencies are obtained by applying an ideal (brick-wall) low-pass filter to $\Omega_X(t)$, $\Omega_Y(t)$ and $\Delta(t)$ in the frequency domain, setting all spectral components above $f_c$ to zero before recomputing the infidelity.
The corresponding values are summarized in Table~\ref{tab:bandwidth}. Figure~\ref{fig:cutoff} shows the resulting infidelity as a function of the cutoff frequency for the Gaussian $\alpha$-RIO pulse after $N=4$ iterations with $\sigma=2\sigma_0$, corresponding to $6.7$~ns for $d/2\pi=-300$~MHz. Figure~\ref{fig:cutoff_vs_sigma} shows more generally how this cutoff frequency decreases with increasing the pulse duration.

\section{GRAPE-Enhanced RDK Pulses}

\begin{figure}[h!]
	\centering
	\includegraphics[width=0.45\textwidth]{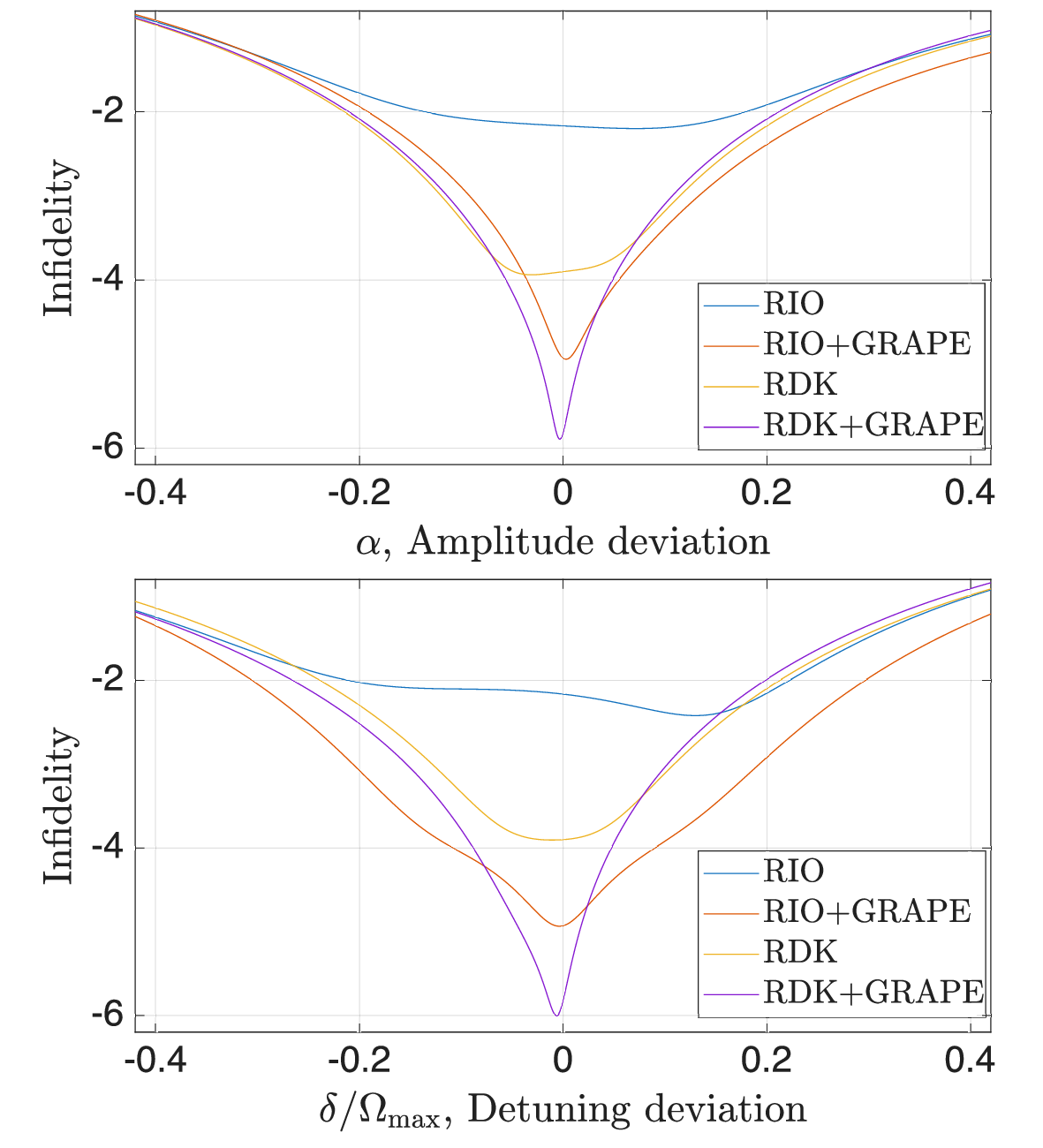}
	\caption{Robustness profiles for the $\alpha\delta$-RIO pulse at $\sigma = 2.5\,\sigma_0$ for the bare $\alpha\delta$-RIO, RIO+GRAPE, RDK ($N=4$), and RDK+GRAPE, the latter achieving the lowest nominal infidelity. Top: amplitude deviation $\alpha$. Bottom: detuning deviation $\delta/\Omega_{\max}$.
	}
	\label{fig:Grape_robustness}
\end{figure}

The GRadient Ascent Pulse Engineering (GRAPE) method is applied to the RIO or RDK pulse (both in the time-reparametrization conditions) as initial condition and denoted RIO+GRAPE and RDK+GRAPE, respectively. To account for leakage, we consider a three-level Hamiltonian. When applying GRAPE to the $\alpha\delta$-RIO pulse, robustness is enforced by augmenting the cost function: specifically, we evaluate the fidelity over a grid of 21 uniformly sampled points $\alpha_k$ in $\alpha\in [-0.05,0.05]$ (for $\delta=0$) and $\delta_k$ in $\delta\in [-0.03,0.03]$ (for $\alpha=0$) (corresponding to $\delta/\Omega_{\mathrm{max}} \in [-0.13,0.13]$), for which we found good results.
We have developed a GRAPE code following Ref.~\cite{JPB-Review}, using backward and forward propagations, an explicit first order derivative, and which takes into account the cost function as the mean infidelity over this set of sampled points,
\begin{equation}
	J = \frac{1}{2}\left(\frac{1}{N_\alpha}\sum_{k=1}^{N_\alpha}|\langle1|\psi_{\alpha_k}\rangle|^2
	+\frac{1}{N_\delta}\sum_{k=1}^{N_\delta}|\langle1|\psi_{\delta_k}\rangle|^2\right),
\end{equation}
where $|\psi_{\alpha_k}\rangle$ and $|\psi_{\delta_k}\rangle$ are the final states obtained under a Hamiltonian with deviations $\alpha = \alpha_k,\, \delta = 0$ and $\alpha = 0,\, \delta = \delta_k$, respectively.

Robustness profiles as a function of both $\alpha$ and $\delta$ are presented for all four scenarios (RIO, RIO+GRAPE, RDK, RDK+GRAPE) in Fig.~\ref{fig:Grape_robustness}. Applying GRAPE on RDK gives a reduction of approximately two orders of magnitude in the nominal infidelity with respect to RDK, and one order of magnitude with respect to RIO+GRAPE.
The pulse duration is taken as $10\sigma$, discretized into $5000$ time steps, and the optimization is initialized once. The gradient correction is weighted by a hyper-Gaussian window $w(t)=\exp[-\tfrac{1}{2}(t/3\sigma)^{10}]$ to suppress oscillations at large times.
In terms of the figure of merit $J$, which is also the GRAPE cost function, for the same optimization parameters RIO+GRAPE converges after 70 iterations to $1-J=10^{-4.18}$, while RDK+GRAPE converges already after 9 iterations to $1-J=10^{-3.73}$. If we restrict the calculation of the cost $J$ over identical intervals $\pm5\%$ for both $\alpha$ and $\delta/\Omega_{\max}$, i.e. $\alpha \in [-0.05,0.05]$ with $\delta=0$, and $\delta/\Omega_{\mathrm{max}} \in [-0.05,0.05]$, we obtain $1-J=10^{-4.5}$ for RDK+GRAPE  and $1-J=10^{-4.4}$ for RIO+GRAPE. The similar values of $J$ are due to the fact that RIO+GRAPE gives a broader robustness profile in $\delta$, albeit shallower, as seen in Fig.~\ref{fig:Grape_robustness}.
We conclude that seeding GRAPE with RDK gives the overall best nominal infidelity and converges eightfold faster.

\begin{figure}[h!]
	\centering
	\includegraphics[width=0.4\textwidth]{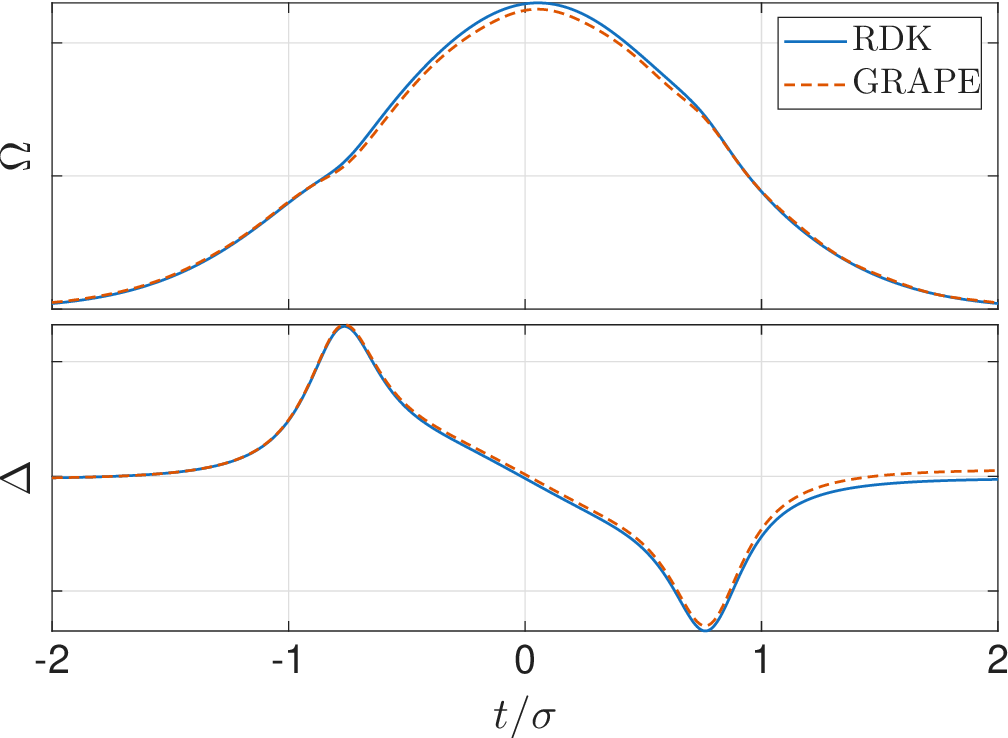}
	\caption{Control fields $\Omega$ and $\Delta$, in a frame where $\Omega$ is real, for the $\alpha\delta$-RIO pulse at $\sigma = 2.5\,\sigma_0$, comparing after 4 DRAG-KAM iterations (RDK) and after enhancing RDK with GRAPE (RDK+GRAPE).}
	\label{fig:grapecontrols}
\end{figure}

The optimized controls $\Omega(t)$ and $\Delta(t)$, obtained in a frame where $\Omega$ is real, are shown for the GRAPE+RDK and RDK-only cases in Fig.~\ref{fig:grapecontrols}. The two pulse shapes are qualitatively similar, indicating that GRAPE introduces only moderate corrections to the RDK solution. We note, that GRAPE introduces oscillations in $\Omega(t)$ at larger times; to assess their physical relevance, we studied the fidelity as a function of the cutoff frequency of a low-pass filter applied to the pulse, and find that the fidelity is nearly identical with and without GRAPE, confirming that the oscillations do not contribute to the population transfer.

\section{Open-system simulation}
\begin{figure}[h!]
	\centering
	\includegraphics[width=0.4\textwidth]{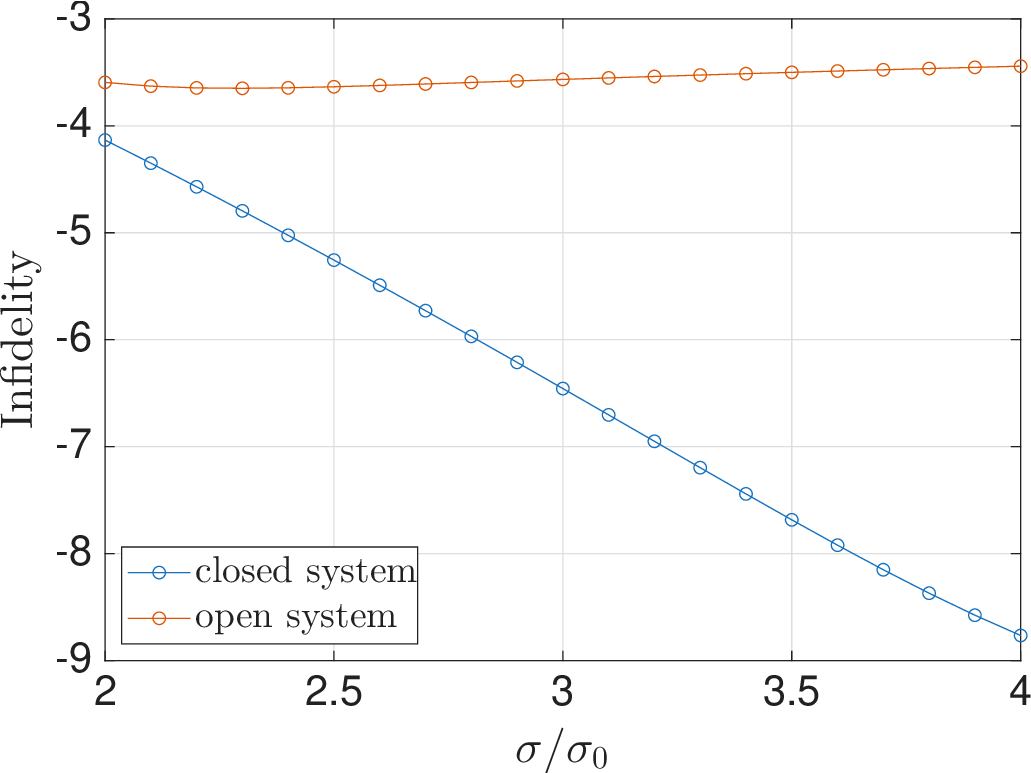}
	\caption{Infidelity as a function of pulse duration $\sigma/\sigma_0$ for the $\alpha$-RIO pulse, for the closed system (blue) and open system (orange), corresponding to $\sigma = 6.7$--$13.3$~ns for $d/2\pi = -300$~MHz.}
	\label{fig:lindblad}
\end{figure}
To assess the impact of decoherence, we simulate the qubit dynamics under a Lindblad master equation
\begin{equation}
	\frac{d\rho}{dt}=-i[H(t),\rho]+\sum_{k}\Big(L_{k}\rho L_{k}^{\dagger}
	-\frac{1}{2}\{\rho,L_{k}^{\dagger}L_{k}\}\Big),
\end{equation}
using the full three-level Hamiltonian $H(t)$ built from the RDK controls $\Omega(t)$ and $\Delta(t)$ after $N=4$ DRAG-KAM iterations. We include two Lindblad operators acting within the qubit subspace $\{|0\rangle,|1\rangle\}$: relaxation, $L_{1}=\sqrt{1/T_1}\,|0\rangle\langle1|$, and pure dephasing, $L_{\phi}=\sqrt{1/(2T_\phi)}\,(|0\rangle\langle0|-|1\rangle\langle1|)$, where $T_\phi$ is related to the standard relaxation and dephasing times $T_1$ and $T_2$ via $1/T_2=1/T_\phi+1/(2T_1)$.
We use standard transmon coherence times $T_1=200\,\mu\text{s}$, $T_2=135\,\mu\text{s}$  and an anharmonicity $d/2\pi=-300$~MHz \cite{Motzoi2024}.

We restrict the Lindblad operators to the qubit subspace $\{|0\rangle,|1\rangle\}$, leaving the transiently populated leakage level $|2\rangle$ decoherence-free; since its relaxation rate is comparable to that of $|1\rangle$ ($\Gamma_{2{\to}1}\approx2\Gamma_{1{\to}0}$ for a transmon) and its population remains small throughout the pulse, this approximation has a negligible effect on the reported infidelities.

The master equation is solved by vectorizing $\rho$, $|\rho\rangle\!\rangle=\sum_{ij}\rho_{ij}|i\rangle|j\rangle$. For arbitrary operators $A,B$ this representation satisfies the identity $|A\rho B\rangle\!\rangle=(A\otimes B^T)|\rho\rangle\!\rangle$, so that the master equation takes the form $\frac{d}{dt}|\rho\rangle\!\rangle=\mathcal{L}\,|\rho\rangle\!\rangle$ with
\begin{align}
	&\mathcal{L}=-i\big(H\otimes\mathbb{I}-\mathbb{I}\otimes H^T\big)\nonumber\\
	&+\sum_k\gamma_k\Big(L_k\otimes L_k^*-\tfrac{1}{2}L_k^\dagger L_k\otimes\mathbb{I}
	-\tfrac{1}{2}\mathbb{I}\otimes(L_k^\dagger L_k)^T\Big),
\end{align}
and is propagated as $|\rho(t+dt)\rangle\!\rangle=e^{\mathcal{L}\,dt}|\rho(t)\rangle\!\rangle$.

Figure~\ref{fig:lindblad} shows the closed- and open-system infidelities as a function of pulse duration $\sigma/\sigma_0$ for the $\alpha$-RIO pulse with standard transmon coherence times $T_1=200\mu$s, $T_2=135\mu$s and anharmonicity $d/2\pi=-300$MHz \cite{Motzoi2024}. For durations reaching at least ${-4}$ closed-system infidelity, the open-system infidelity remains around ${-3.65}$ to ${-3.45}$. Similar results are obtained for the $\alpha\delta$-RIO pulse, with the range ${-3.55}$ to ${-3.45}$. Applying GRAPE prior to the Lindblad propagation gives infidelities in the same range, with negligible improvement, if any.


\begin{thebibliography}{99}
	\bibitem{Nielsen} M.A. Nielsen and I.L. Chuang, Quantum Computation
	and Quantum Information (Cambridge University Press,
	Cambridge, U.K., 2000).
	\bibitem{Koch} C.P. Koch, U. Boscain, T. Calarco, G. Dirr, S. Filipp, S.J. Glaser, R. Kosloff, S. Montangero, T. Schulte-Herbr\"uggen, D. Sugny, F.K. Wilhelm, Quantum optimal control in quantum technologies. Strategic report on current status, visions and goals for research in Europe, EPJ Quantum Technol. \textbf{9}, 19 (2022).
	
	\bibitem{SWAP} N. Sangouard, X. Lacour, S. Gu{\'e}rin, H.R. Jauslin, Fast SWAP gate by adiabatic passage, Phys. Rev. A \textbf{72}, 062309 (2005).
	\bibitem{Vitanov_Raman} N.V. Vitanov, A.A. Rangelov, B.W. Shore, and K. Bergmann, Stimulated Raman adiabatic passage in physics, chemistry, and beyond, Rev. Mod. Phys. \textbf{89}, 015006 (2017).
	\bibitem{Sorin} M. Kuzmanovic, I. Bj\"orkman, J.J. McCord, S. Dogra, and G.S. Paraoanu, High-fidelity robust qubit control by phase-modulated pulses, Phys. Rev. Res. \textbf{6}, 013188 (2024).
	
	
	\bibitem{STA} D. Gu\'ery-Odelin, A. Ruschhaupt, A. Kiely, E. Torrontegui,
	S. Mart\'inez-Garaot, and J. G. Muga, Shortcuts to adiabaticity:
	Concepts, methods, and applications, Rev. Mod. Phys. \textbf{91},
	045001 (2019).
	
	\bibitem{robustNJP} A Ruschhaupt, X. Chen, D. Alonso and J. G Muga, Optimally robust shortcuts to population inversion in two-level quantum systems, N. J. Phys. \textbf{14}, 093040 (2012).
	\bibitem{SSSP}  D. Daems, A. Ruschhaupt, D. Sugny, and S. Gu\'erin, Robust Quantum Control by a Single-Shot Shaped Pulse, Phys. Rev. Lett. \textbf{111}, 050404 (2013).
	\bibitem{Clerk} F. Setiawan, P. Groszkowski, H. Ribeiro, and A.A. Clerk, Analytic Design of Accelerated Adiabatic Gates in Realistic Qubits: General
	Theory and Applications to Superconducting Circuits, PRX Quantum \textbf{2}, 030306 (2021).
	\bibitem{STAAQC} N. N. Hegade, K. Paul, Y. Ding, M. Sanz, F. Albarr{\'a}n-Arriagada, E. Solano, and X. Chen, Shortcuts to Adiabaticity in Digitized Adiabatic Quantum Computing, Phys. Rev. Appl. 15, 024038 (2021).
	
	
	\bibitem{Wimperis} S. Wimperis, Broadband, narrowband, and passband composite pulses for use in advanced NMR experiments, J. Magn. Reson. \textbf{109}, 221 (1994).
	\bibitem{Cummins} H.K. Cummins, G. Llewellyn, and J.A. Jones, Tackling systematic errors in quantum logic gates with composite rotations, Phys. Rev. A \textbf{67}, 042308 (2003).
	\bibitem{Torosov} B.T. Torosov, S. Gu{\'e}rin, and N.V. Vitanov, High-fidelity adiabatic passage by composite sequences of chirped pulses, Phys. Rev. Lett. \textbf{106}, 233001 (2011).
	\bibitem{Jones}  J.A. Jones, Designing short robust NOT gates for quantum computation, Phys. Rev. A \textbf{87}, 052317 (2013).
	\bibitem{Torosov2} B.T. Torosov and N.V. Vitanov, Arbitrarily accurate variable rotations on the Bloch sphere by composite pulse sequences, Phys. Rev. A \textbf{99}, 013402 (2019).
	\bibitem{Gevorgyan}  H.L. Gevorgyan and N.V. Vitanov, Ultrahigh-fidelity
	composite rotational quantum gates, Phys. Rev. A \textbf{104}, 012609 (2021).
	
	\bibitem{Barnes} J. Zeng, C.H. Yang, A. S. Dzurak, and E. Barnes, Geometric formalism for constructing arbitrary single-qubit dynamically corrected gates, Phys. Rev. A \textbf{99}, 052321 (2019).
	
	\bibitem{Dridi} G. Dridi, K. Liu, S. Gu{\'e}rin, Optimal robust quantum
	control by inverse geometric optimization, Phys. Rev.
	Lett. \textbf{125}, 250403 (2020).
	
	\bibitem{Laforgue}X. Laforgue, G. Dridi and S. Gu{\'e}rin, Optimal robust
	stimulated Raman exact passage by inverse optimization,
	Phys. Rev. A \textbf{105}, 032807 (2022).
	
	\bibitem{Laforgue2}X. Laforgue, G. Dridi, and S. Gu{\'e}rin, Optimal robust
	quantum control against pulse inhomogeneities: Analytic
	solutions, Phys. Rev. A \textbf{106}, 052608 (2022).
	\bibitem{Motzoi2009}
	F. Motzoi, J.M. Gambetta, P. Rebentrost, and F.K. Wilhelm, Simple pulses for elimination of leakage in weakly nonlinear qubits,
	Phys. Rev. Lett. \textbf{103}, 110501 (2009).
	
	\bibitem{Gambetta2011}
	J.M. Gambetta, F. Motzoi, S.T. Merkel, and F.K. Wilhelm, Analytic control methods for high-fidelity unitary operations in a weakly nonlinear oscillator,
	Phys. Rev. A \textbf{83}, 012308 (2011).
	
	\bibitem{Theis2018}
	L.S. Theis, F. Motzoi, S. Machnes, and F.K. Wilhelm, Counteracting systems of diabaticities using DRAG controls:
	The status after 10 years,
	EPL \textbf{123}, 60001 (2018).
	
	\bibitem{Werninghaus2021}
	M. Werninghaus, D.J. Egger, F. Roy, S. Machnes, F.K. Wilhelm, and S. Filipp, Leakage reduction in fast superconducting qubit gates via optimal control, 
	npj Quantum Information \textbf{7}, 14 (2021).
	
	\bibitem{Tonchev}
	H.G. Tonchev, B.T. Torosov and N.V. Vitanov,  Robust, fast, and high-fidelity composite single-qubit gates for superconducting transmon qubits, Phys. Rev. A \textbf{112}, 012605 (2025).
	
	\bibitem{YongHai} Y.-J. Hai, J. Li, J. Zeng, D. Yu, and X.-H. Deng, Geometric correspondence of noisy quantum dynamics and universal robust quantum
	gates, arXiv: 2210.14521v5.
	\bibitem{KangyuanYi} K. Yi , Y.-J. Hai, K. Luo, J. Chu, L. Zhang, Y. Zhou, Y. Song, S. Liu, T. Yan, X.-H. Deng, Y. Chen, and D. Yu, Robust Quantum Gates against Correlated Noise in Integrated Quantum Chips, Phys. Rev. Lett. \textbf{132}, 250604 (2024).
	\bibitem{Scherer-II-97}
	W. Scherer,
	Quantum averaging II: Kolmogorov's algorithm,
	J. Phys. A {\bf 30}, 2825 (1997).
	\bibitem{Scherer-prl95}
	W. Scherer,
	Superconvergent perturbation method in quantum mechanics,
	Phys. Rev. Lett. {\bf 74}, 1495 (1995).
	\bibitem{PhysicA}  M. Amniat-Talab S. Gu\'{e}rin and H.R. Jauslin, 
	Quantum averaging and resonances: Two-level atom
	in a one-mode quantized field, J. Math. Phys. {\bf 46}, 042311 (2005).
	\bibitem{Review}
	S. Gu\'{e}rin and H.R. Jauslin, Control of
	quantum dynamics by laser pulses -- Adiabatic Floquet
	theory, Adv. Chem. Phys {\bf 125}, 147 (2003).
	
	\bibitem{Bender} C.M. Bender and S. A. Orszag, Advanced Mathematical Methods for Scientists and Engineers I: Asymptotic Methods and Perturbation Theory (Springer-Verlag, New York, 1999).
	
	\bibitem{BerryHyper} M.V. Berry and C.J. Howls, Hyperasymptotics, Proc. Roy. Soc. London A \textbf{430}, 653 (1990).
	
	\bibitem{Boyd} J.P. Boyd, The Devil's Invention: Asymptotic, Superasymptotic and Hyperasymptotic Series, Acta Applicandae Mathematicae \textbf{56}, 1 (1999).
	
	\bibitem{BerrySA} M. V. Berry, Histories of adiabatic quantum transitions, Proc. R. Soc. London A \textbf{429}, 61 (1990).
	
	\bibitem{Joye} A Joye, Non-trivial prefactors in adiabatic transition probabilities induced by high-order complex degeneracies, J. Phys. A \textbf{26} 6517 (1993)
	
	\bibitem{Holthaus} K. Drese and M. Holthaus, Perturbative and nonperturbative processes in adiabatic population transfer, Eur. Phys. J. D \textbf{3}, 73 (1998).
	
	\bibitem{code-data} https://github.com/meri-harutyunyan/Iterative-DRAG-KAM-Algorithm
	
	\bibitem{Unanyan} R.G. Unanyan, L.P. Yatsenko, K. Bergmann, B.W. Shore, Laser-induced adiabatic atomic reorientation with control of diabatic losses. Opt. Commun. \textbf{139}, 48 (1997).
	\bibitem{Berry} M.V. Berry, Transitionless quantum driving. J. Phys. A Math. Theor. \textbf{42}, 365303 (2009).
	
	\bibitem{XiChen} S. Ib\`a\~nez, Xi Chen, and J. G. Muga, Improving shortcuts to adiabaticity by iterative interaction pictures, Phys. Rev. A \textbf{87}, 043402 (2013).
	\bibitem{SM} See Supplemental Material at [URL], which includes additional details about the algorithm, the analysis of the pulse on spectral properties, further improvement by GRAPE, and the effect of decoherence.
	
	\bibitem{ZhuJPB}
	J. Zhu, X. Laforgue, X. Chen, S. Gu{\'e}rin, Robust quantum control by smooth quasi-square pulses, J. Phys. B \textbf{55}, 194001 (2022).
	\bibitem{Motzoi2024} B. Li, T. Calarco, and F. Motzoi, Experimental error suppression in cross-resonance gates via multi-derivative pulse shaping, npj Quantum Inf \textbf{10}, 66 ( 2024).
	\bibitem{QuditMotzoi} B. Li, F.A. C{\'a}rdenas-L{\'o}pez, A. Lupascu, and F. Motzoi, Universal pulses for superconducting qudit ladder gates, PRX Quantum \textbf{6}, 030357 (2025).
	\bibitem{Qudit}  B. Rousseaux, S. Gu{\'e}rin, and N.V. Vitanov, Arbitrary qudit gates by adiabatic passage, Phys. Rev. A \textbf{87}, 032328 (2013).
	\bibitem{FeiYan} R. Wang, Y. Feng, Y. Zhang, J. Ding, B. Li, F. Motzoi, Y. Gao, H. Xu, Z. Yang, W. Nuerbolati, H. Yu, W. Sun, and F. Yan, Suppressing Spurious Transitions Using Spectrally Balanced Pulse, Phys. Rev. Lett. \textbf{135}, 160804 (2025).
	
	\bibitem{JPB-Review} Q. Ansel, E. Dionis, F. Arrouas, B. Peaudecerf, S. Gu{\'e}rin, D. Gu{\'e}ry-Odelin and D. Sugny,  Introduction to Theoretical and Experimental aspects of Quantum Optimal Control, J. Phys. B  \textbf{57}, 133001 (2024).
	
	\bibitem{Guery-Odelin} N. Dupont, G. Chatelain, L. Gabardos , M. Arnal, J. Billy, B. Peaudecerf , D. Sugny, and D. Gu{\'e}ry-Odelin, Quantum State Control of a Bose-Einstein Condensate in an Optical Lattice, PRX Quantum \textbf{2}, 040303 (2021).
	
	\bibitem{Kiely} P.M. Poggi, G. De Chiara, S. Campbell, and A. Kiely, Universally Robust Quantum Control, Phys. Rev. Lett. {\bf 132}, 193801 (2024).
	\bibitem{Deng2025} Y.-J. Hai, Y. Song, J. Li, J. Zeng, and X.-H. Deng, Geometric correspondence of noisy quantum dynamics and universal robust quantum gates, Phys. Rev. App. \textbf{23}, 054002 (2025).
	
\end{thebibliography}

\vfill

\begin{thebibliography}{99}
	\bibitem{QuditMotzoi} B. Li, F.A. C{\'a}rdenas-L{\'o}pez, A. Lupascu, and F. Motzoi, Universal pulses for superconducting qudit ladder gates, PRX Quantum \textbf{6}, 030357 (2025).
	\bibitem{Qudit}  B. Rousseaux, S. Gu{\'e}rin, and N.V. Vitanov, Arbitrary qudit gates by adiabatic passage, Phys. Rev. A \textbf{87}, 032328 (2013).
	\bibitem{Motzoi2024} B. Li, T. Calarco, and F. Motzoi, Experimental error suppression in cross-resonance gates via multi-derivative pulse shaping, npj Quantum Inf \textbf{10}, 66 ( 2024).
	\bibitem{Dridi} G. Dridi, K. Liu, S. Gu{\'e}rin, Optimal robust quantum
	control by inverse geometric optimization, Phys. Rev.
	Lett. \textbf{125}, 250403 (2020).
	\bibitem{ZhuJPB}
	J. Zhu, X. Laforgue, X. Chen, S. Gu{\'e}rin, Robust quantum control by smooth quasi-square pulses, J. Phys. B \textbf{55}, 194001 (2022).
	\bibitem{JPB-Review} Q. Ansel, E. Dionis, F. Arrouas, B. Peaudecerf, S. Gu{\'e}rin, D. Gu{\'e}ry-Odelin and D. Sugny,  Introduction to Theoretical and Experimental aspects of Quantum Optimal Control, J. Phys. B  \textbf{57}, 133001 (2024).
	
\end{thebibliography}
\end{document}